\documentclass{aa}
\usepackage[T1]{fontenc}
\usepackage{color}
\usepackage{amstext}
\usepackage{amssymb}
\usepackage{graphicx}

\makeatletter

\providecommand{\tabularnewline}{\\}

\usepackage[varg]{txfonts}
\bibpunct{(}{)}{;}{a}{}{,}

\makeatother

\usepackage{babel}

\begin{document}

\title{Sustained oscillations in interstellar chemistry models}
\author{Evelyne Roueff\inst{1}\and Jacques Le Bourlot\inst{1,2}}
\institute{LERMA, Observatoire de Paris, PSL Research University, CNRS, Sorbonne
Universit\'e, UPMC Univ. Paris 06, F-92190, Meudon, France \email{evelyne.roueff@obspm.fr}
\and Universit\'e de Paris, Paris, France \email{jacques.lebourlot@obspm.fr}
}

\date{Received ...; Accepted...}
\abstract{Nonlinear behavior in interstellar chemical models has been recognized
for 25 years now. Different mechanisms account for the possibility
of multiple fixed-points at steady-state, characterized by the ionization
degree of the gas.}{Chemical oscillations are also a natural behavior
of nonlinear chemical models. We study under which conditions spontaneous
sustained chemical oscillations are possible, and what kind of bifurcations
lead to, or quench, the occurrence of such oscillations.}{The well-known ordinary differential equations (ODE) integrator VODE
was used to explore initial conditions and parameter space in a gas
phase chemical model of a dark interstellar cloud.}{We recall that the time evolution of the various
chemical abundances under fixed temperature conditions depends on
the density over cosmic ionization rate $n_{\mathrm{H}}/\zeta$ ratio.
We also report the occurrence of naturally sustained oscillations
for a limited but well defined range of control parameters. The period
of oscillations is within the range of characteristic timescales
of interstellar processes and could lead to spectacular resonances
in time-dependent models. Reservoir species ($\mathrm{C}$, $\mathrm{CO}$, $\mathrm{NH_3}$, ...)
oscillation amplitudes are generally less than a factor two. However,
these amplitudes reach a factor ten to thousand for low abundance species, e.g. $\mathrm{HCN}$, $\mathrm{ND_3}$,
that may play a key role for diagnostic purposes.
The mechanism
responsible for oscillations is tightly linked to the chemistry of
nitrogen, and requires long chains of reactions such as found in multi-deuteration
processes.}{}
\maketitle

\section{Introduction\label{sec:Introduction}}

The evolution of interstellar chemical abundances through formation/destruction
chemical processes is described by a set of nonlinear differential
equations that fall under the frame of dynamical systems. Then, it
is not surprising to find some of the standard features characteristic
of these systems. In particular multiple fixed-points for a given
set of control parameters have been demonstrated by \citet{1992MNRAS.258P..45P}
and first analyzed by \citet{1993ApJ...416L..87L,1995A&A...297..251L,1995A&A...302..870L}.
The two stable solutions correspond to very different chemical signatures
and have been labelled as LIP, corresponding to a Low Ionization Phase
and HIP for an High Ionization Phase. LIP features are described by
abundant saturated molecules and molecular ions whereas HIP signatures
rather point to abundant radicals, carbon chains and atomic ions.
Although these results led to some debate (e.g. \citealt{1995A&A...296..779S}),
they were later confirmed by various groups (\citealt{1998A&A...334.1047L,2003A&A...399..583C,2006ApJ...645..314B,2019ApJ...887...67D}).
Some puzzling observations were even suggested to be representative
of the HIP phase by, e.g., \citet{1997A&A...318..579G} or \citet{2011ApJ...740L...4C}.

However, apart from a passing mention in \citet{1995A&A...297..251L},
no steady chemical oscillations have been reported to date. This is
rather surprising, as such a behavior is a normal feature of nonlinear
sets of chemical equations (see \citealt{gray1994chemical}) and results
probably from the limited range of integration time and parameter
values considered in the current time-dependent chemical studies applied
to interstellar clouds (\citealt{2015ApJS..217...20W}). Various time
dependent behaviors including oscillations are explicitly quoted in
\citet{2018ApJ...868...41M}. This paper will be discussed below,
as we believe their study to be not fully complete. Furthermore, our
analysis explains naturally a large fraction of their findings.

In this paper, we describe in some details naturally occurring oscillations
in a model of dark cloud chemistry. The underlying formalism is presented
in Sect.~\ref{sec:Model}, a fiducial oscillatory solution is introduced
in Sect.~\ref{sec:Results}. We discuss the effect of various control
parameters in Sect.~\ref{sec:Variation-of-control} and present our
analysis and a discussion in Sect.~\ref{sec:Discussion}.

\section{Time dependent chemical equations\label{sec:Model}}

We restrict ourself to an idealized dark cloud. That is, photo processes
come only from secondary photons following the creation of high energy
electrons by cosmic-ray. These electrons excite the different electronic
states of $\mathrm{H}_{2}$ that eventually decay by emitting ultraviolet
photons in the $750-1750\,\textrm{\AA}$ window (\citealt{1989ApJ...347..289G}).

We also make two very strong restrictions by assuming a constant temperature
and discarding gas-grain chemical interactions. We are fully aware
that, as a consequence, the resulting models cannot be readily applied
to ''real'' astrophysical objects, nor be compared to observations.
But this is a necessary preliminary step to try and understand the
chemical mechanisms at work. Extended models that overcome one or
both of these approximations will be introduced later (note however
a brief discussion in the last section).

On the other hand, our chemical model includes an extended and detailed
chemical network including 254 species, coupled by 4872 reactions
taken from our previous studies (\citealt{2015A&A...576A..99R,2018ApJ...853L..22C}).
In particular it includes a detailed deuteration scheme with multiple
deuterated species from $\mathrm{NH}_{3}$ to $\mathrm{ND}_{3}$ (\citealt{2005A&A...438..585R}),
$\mathrm{H_{2}CO}$ to $\mathrm{D_{2}CO}$, $\mathrm{H_{2}CS}$ to
$\mathrm{D_{2}CS}$ (\citealt{2005ApJ...620..308M}), $\mathrm{CH_{4}}$
to $\mathrm{CD_{4}}$ . We introduce the value of the ortho-to-para ratio of molecular
hydrogen, $(O/P)$, which drives the deuteration efficiency through
the $\mathrm{H}_{3}^{+}+\mathrm{HD}\rightleftarrows\mathrm{H}_{2}\mathrm{D}^{+}+\mathrm{H}_{2}$
reaction (\citealt{1992A&A...258..479P}) as an external fixed parameter,
following \citet{2015A&A...576A..99R}. This parameter is also critical
for the initial step of the nitrogen hydrogenation reaction $\mathrm{N}^{+}+\mathrm{H}_{2}\rightarrow\mathrm{NH^{+}}+\mathrm{H}$
which is slightly endothermic when $\mathrm{H}_{2}$ is in its para
form as first emphasized by \citet{1991A&A...242..235L}. We follow
the prescription of \citet{dislaire:12} based on the experimental
studies of \citet{1988JChPh..89.2041M} to account for this reaction.
We also include the experimental reaction rate coefficients to account
for $\mathrm{N}^{+}+\mathrm{HD}$ and $\mathrm{N}^{+}+\mathrm{D}_{2}$
reactions.

Parameters used for various models are displayed in Table~\ref{tab:Control-parameters.}.
The range shown is typical of the values explored, although we did
not compute all possible models.
\footnote{The code is made available at http://ism.obspm.fr.}

\begin{table*}
\caption{Control parameters.\label{tab:Control-parameters.}}

\centering%
\begin{tabular}{cccc}
\hline
\hline Parameter & Unit & Range & Description\tabularnewline
\hline
$n_{\mathrm{H}}$ & $\mathrm{cm}^{-3}$ & $10^{3}-10^{7}$ & Hydrogen nucleus number density\tabularnewline
$\zeta$ & $10^{-17}\,\mathrm{s}^{-1}$ & $1-300$ & Hydrogen cosmic-ray ionization rate\tabularnewline
$T$ & $\mathrm{K}$ & $5-20$ & Gas temperature\tabularnewline
$\delta_{\mathrm{C}}$ & - & $10^{-5}-10^{-4}$ & Carbon gas-phase abundance\tabularnewline
$\delta_{\mathrm{N}}$ & - & $10^{-5}-10^{-4}$ & Nitrogen gas-phase abundance\tabularnewline
$\delta_{\mathrm{O}}$ & - & $10^{-5}-10^{-4}$ & Oxygen gas-phase abundance\tabularnewline
$\delta_{\mathrm{S}}$ & - & $8\,10^{-8}$ & Sulfur gas-phase abundance\tabularnewline
\hline
\end{tabular}

\end{table*}

\subsection{Nondimensional equations}

One major advantage of using a constant temperature is that it leads
to constant chemical reaction rates once the other external parameters
are chosen. In a low density gas only two body collisions are possible.
So the time evolution of any species $\mathrm{X}_{i}$ abundance $n_{i}=n\left(\mathrm{X}_{i}\right)$
writes:
\begin{equation}
\frac{dn_{i}}{dt}=\sum_{l,m}k_{l,m}^{i}\left(T\right)\,n_{l}\,n_{m}+\sum_{j}k_{j}^{i}\left(T\right)\,n_{j}\label{eq:01} .
\end{equation}

Two bodies rates $k_{l,m}^{i}$ are in $\mathrm{cm}^{3}\,\mathrm{s}^{-1}$
and one body rates $k_{j}^{i}$ are in $\mathrm{s}^{-1}$. These condensed
equations account both for formation/destruction processes with appropriate
signs. Given our hypothesis, all $k_{j}^{i}$ are proportional to
the molecular hydrogen cosmic-ray ionization rate $\zeta$. This
suggests to use $1/\zeta$ as a characteristic timescale, and to
introduce a nondimensional time variable $\tau=\zeta\,t$. Writing
now $n_{i}=n_{\mathrm{H}}\,x_{i}$, with $x_{i}$ the relative abundance
of species $\mathrm{X}_{i}$, we can introduce nondimensional rates
for all chemical reactions by writing:
$k_{l,m}^{i}=\frac{\zeta}{n_{\mathrm{H}}}\,r_{l,m}^{i}\,;\quad k_{j}^{i}=\zeta\,q_{j}^{i}\,$.

Subsequently, Eq.~(\ref{eq:01}) reduces to:
\begin{equation}
\frac{dx_{i}}{d\tau}=\sum_{l,m}r_{l,m}^{i}\,x_{l}\,x_{m}+\sum_{j}q_{j}^{i}\,x_{j}\,.\label{eq:02}
\end{equation}

This analysis shows that we can lower the number of control parameters
by defining an ``ionization efficiency parameter'' $I_{ep}=\,{\displaystyle \frac{n_{\mathrm{H}}}{10^{4}\,\mathrm{cm}^{-3}}}\,{\displaystyle \frac{10^{-16}\,\mathrm{s}^{-1}}{\zeta}}$
or $I_{ep}=10^{-20}\,{\displaystyle \frac{n_{\mathrm{H}}}{\zeta}}$
(unitless), which leads to values of $I_{ep}$ from $1$ to $10$
for typical interstellar conditions. This property has been previously recognized for steady-state
results (\citealt{1996A&A...306L..21L,2006ApJ...645..314B}) or for
the maxima corresponding to so-called early-time results (\citealt{1998A&A...334.1047L}).
However, to our knowledge, this occurrence was never taken full advantage of
for interstellar applications. Let us also point out that this property
is maintained for gas--grain interactions as long as a single grain
temperature is considered and a standard rate equations formalism
is used to describe surface chemistry (\citealt{1992ApJS...82..167H}).

\subsection{Dynamical system properties}

Equation~(\ref{eq:02}) is in the standard form of a dynamical system:
\begin{equation}
\dot{X}=F\left(X,\alpha\right)\,,
\end{equation}
where $X$ is the vector of (unknown) variables, $\alpha$ a set of
(fixed) control parameters, and $F$ a vector of functions. Below, we summarize
the main properties of dynamical systems as described in the
literature; for example \citet{gray1994chemical}. Fixed points of this system
are found by setting $F\left(X,\alpha\right)=0$. The (possibly multiple)
solutions may or may not be stable. The stability can be assessed
by computing the eigenvalues of the Jacobian matrix $J$ at the fixed
point. If $\bar{X}$ is a fixed point, then:
\begin{equation}
J\left(\bar{X}\right)=\left(\left.\frac{\partial F_{i}}{\partial X_{j}}\right|_{\bar{X}}\right)\,.
\end{equation}

Close to $\bar{X}$, the system~(\ref{eq:02}) can be linearized.
Its evolution is then given by
\begin{equation}
X\left(\tau\right)=\bar{X}+\sum_{i}A_{i}\,\exp\left(\lambda_{i}\,\tau\right)\,,
\end{equation}
where $\lambda_{i}$ are the eigenvalues of $J\left(\bar{X}\right)$.
We see that, if all real parts $\Re\left(\lambda_{i}\right)<0$, then
the fixed point is stable, and evolution may lead to it, provided
that all control parameters remain constant during the evolution.
If a single $\lambda_{i}$ has a positive real part, then the fixed
point is unstable, and the system leaves its vicinity. As soon as
$X\left(\tau\right)-\bar{X}$ is no longer infinitely small, the
linearized approximation does not apply and higher order terms control
the evolution.

This is independent of whether $\lambda_{i}$ eigenvalues
have an imaginary part or not. If they do, then they occur in pairs
of complex conjugates. Writing $\lambda_{i}=\mu_{i}+i\,\omega_{i}$,
all $X\left(\tau\right)$ include a contribution of the form
$\exp\left(\mu_{i}\,\tau\right)\,\cos\left(\omega_{i}\,\tau+\phi_{i}\right)$,
which is an oscillatory contribution.
\begin{itemize}
\item If the fixed point is stable, then the system behaves as a damped
oscillator, with a damping characteristic time of $\tau_{d}={\displaystyle \frac{1}{\mu_{j}}}$,
where $\mu_{j}$ is the smallest eigenvalue (in absolute value).
If the associated $\lambda_{i}$ has an imaginary part, then there
is an oscillatory damping.
\item If the fixed point is unstable, the system leaves its vicinity with
a timescale associated to the most positive eigenvalue. Here again,
an oscillatory component may or may not be present.
\end{itemize}
The subsequent evolution depends on the topology of the phase space.
Evolution may lead to another (stable) fixed point (steady-state),
or to sustained oscillations if a stable limit cycle exists. In the
following, we show how a stable fixed point undergoes a Hopf bifurcation,
giving birth to a stable limit cycle at a specific value of a control
parameter. This occurs when the real part of an eigenvalue of the
Jacobian matrix crosses a value of zero. We note that this also explains
the phenomenon of ``critical slow-down'' observed in the vicinity
of a bifurcation point. As the evolution depends on $\exp\left(\mu_{i}\,\tau\right)$,
the damping or growing time diverges as $\mu_{i}\rightarrow0$. An
example is given below in Sect.~\ref{subsec:Effect-of-temperature}.

\subsection{The role of initial conditions}

If several attractors exist (e.g., one or more stable fixed points,
and a stable limit cycle) then the evolution depends critically on
the chosen initial conditions. Different initial conditions may lead
to one or the other attractor as previously shown in \citet{1993ApJ...416L..87L}.
The initial part of the evolution has thus no particular meaning,
depending solely on an arbitrary choice of initial conditions. Relaxation
to the attractor depends mainly on its stability character and its
distance \textendash-{} in parameter space \textendash-{} from a possible
bifurcation point. We would like to stress here that the
so-called early-time behavior is entirely dependent on an arbitrary
choice of initial conditions; it should be considered with great care
as it provides almost no information on the intrinsic behavior of
the system.

When application to interstellar medium conditions is examined, one
should remember that interstellar clouds do not form out of nothing
at a given time, but rather result from fluctuations within a turbulent
medium that is permanently stirred by various dynamical process (e.g.,
supernovae or the differential rotation of the galaxy). Part of the
gas disappears in new stars and fresh material is constantly added
to the medium by the evolution of stars (stellar winds, loss of envelopes,
supernova) or cooling flows from intergalactic material.

Thus, when a dense cloud separates from its environment, its composition
is typical of the diffuse medium and quite different from the arbitrary
simple values often found in the literature (e.g., $\mathrm{H}_{2}$,
$\mathrm{C}^{+}$, $\mathrm{O}$, $\mathrm{N}$, ...). Model results
obtained with such initial conditions may therefore be significantly misleading.

\section{The oscillatory fiducial model\label{sec:Results}}

We report and discuss now our finding of oscillatory solutions obtained
through a mixture of serendipity and systematic exploration of the
model parameters displayed in Table~\ref{tab:Control-parameters.}.

\subsection{Typical oscillations}

Table~\ref{tab:Reference-model-parameters.} shows the values of
the physical parameters leading to an oscillatory behavior of the
reduced abundances, and defines our fiducial model.

\begin{table}[h]
\caption{Reference model parameters
$\left(I_{ep}={\displaystyle \frac{n_{\mathrm{H}}}{10^4 cm^{-3}} \cdot \frac{10^{-16}\, s^{-1}}{\zeta} } \right)$.
\label{tab:Reference-model-parameters.}}
\centering
\begin{tabular}{ccccccc}
\hline
\hline $I_{ep}$ & $T$ & $\delta_{\mathrm{C}}$ & $\delta_{\mathrm{N}}$ & $\delta_{\mathrm{O}}$ & $\delta_{\mathrm{S}}$ & $\delta_{\mathrm{Fe}}$\tabularnewline
\hline
$4$ & $11\,\mathrm{K}$ & $3\,10^{-5}$ & $6\,10^{-5}$ & $3\,10^{-5}$ & $8\,10^{-8}$ & $1.5\,10^{-8}$\tabularnewline
\hline
\end{tabular}
\end{table}

\begin{figure*}
\centering
\includegraphics[width=1\columnwidth]{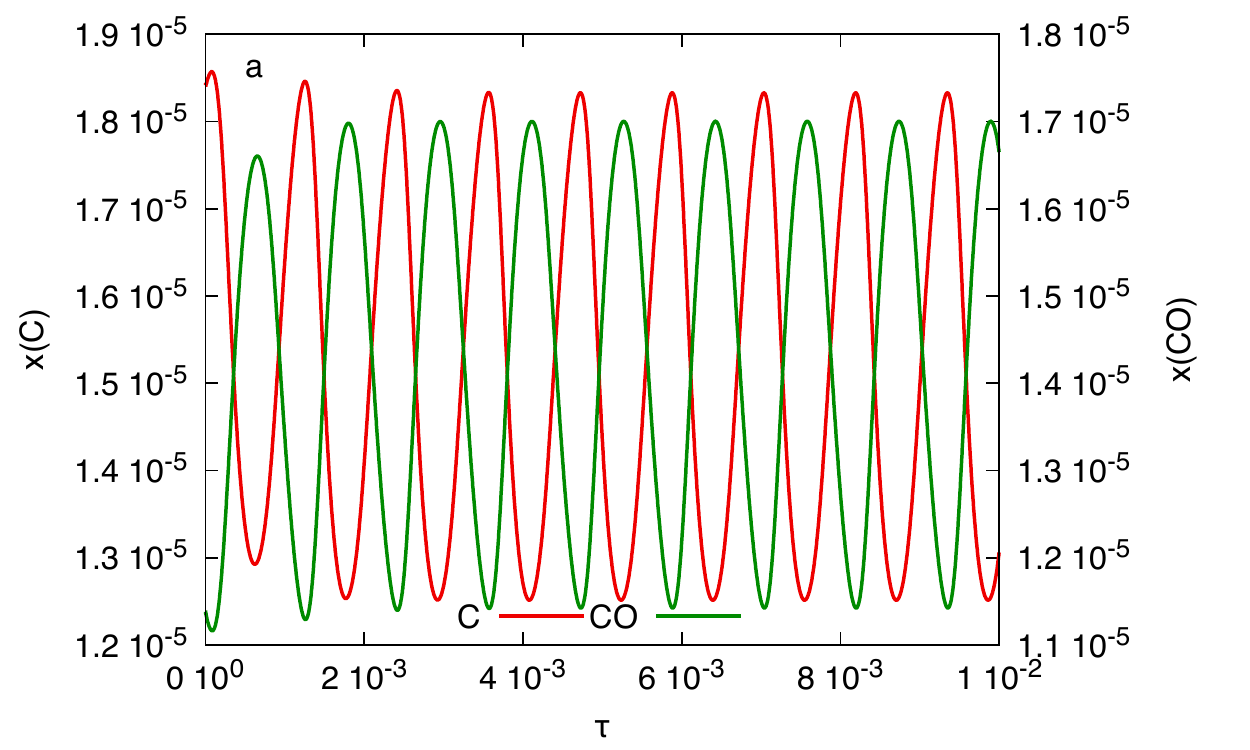}
\includegraphics[width=1\columnwidth]{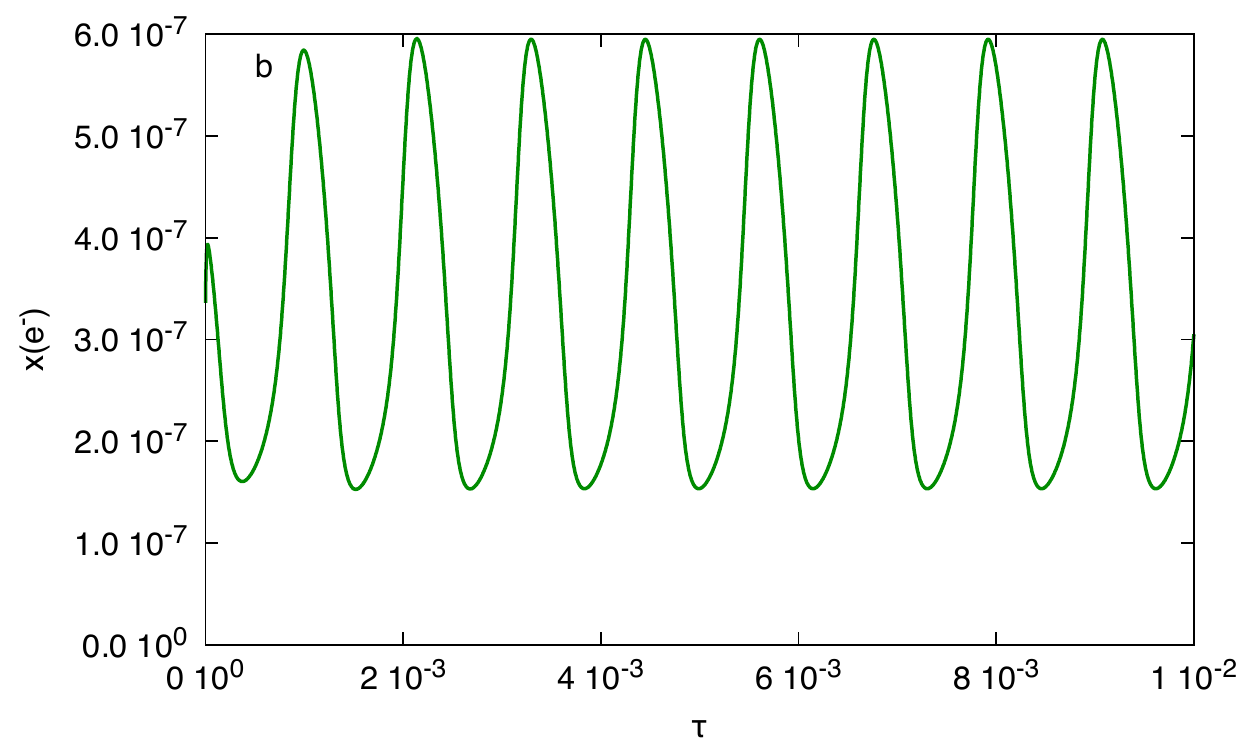}
\includegraphics[width=1\columnwidth]{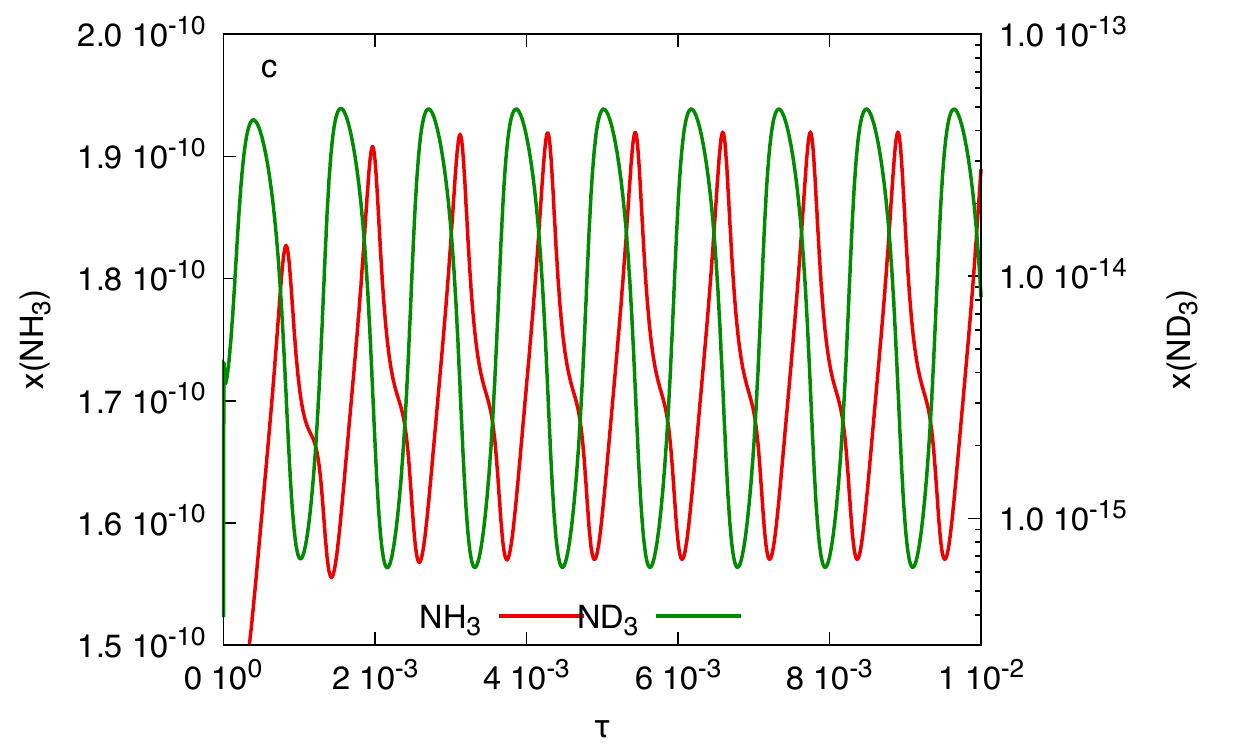}
\includegraphics[width=1\columnwidth]{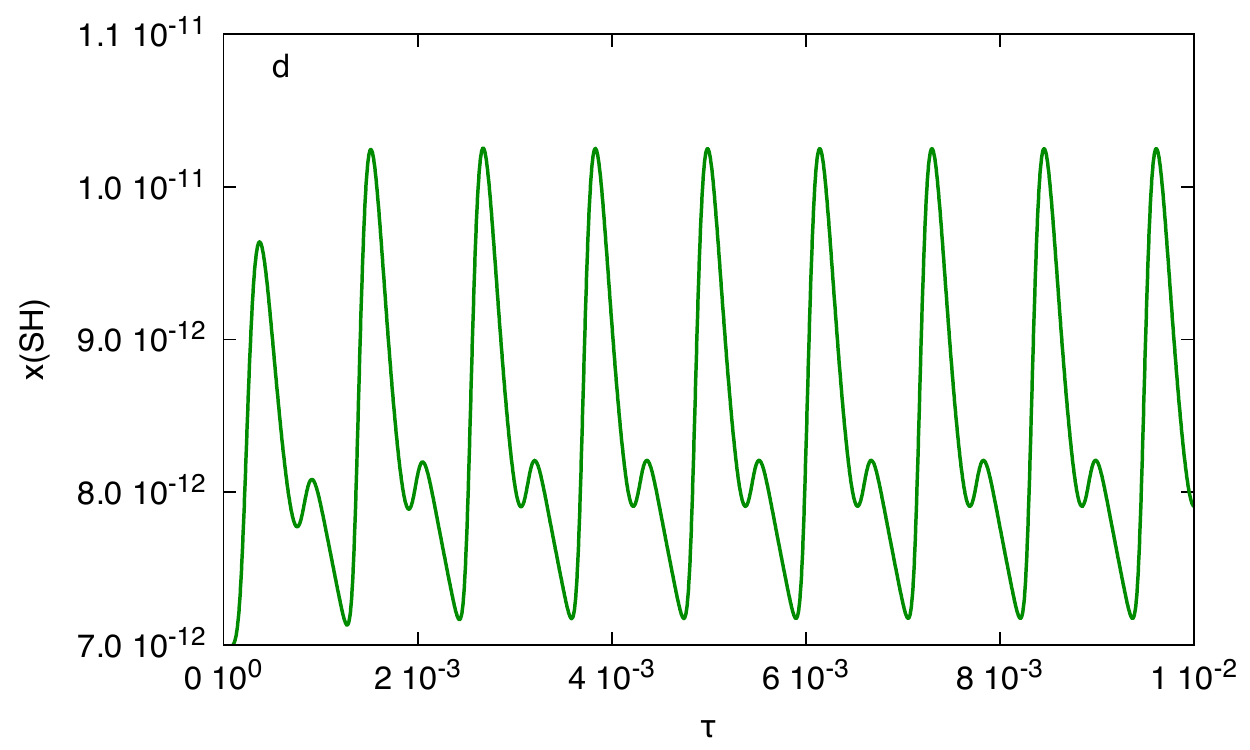}
\caption{Typical oscillations obtained in the fiducial model. $\tau=10^{-3}$
corresponds to $\sim0.634\,\mathrm{Myr}$ for $\zeta=5\,10^{-17}\,\mathrm{s}^{-1}$.\label{fig:Typical-oscilations.}}
\end{figure*}

Examples of oscillations are shown in Fig.~\ref{fig:Typical-oscilations.}.

All quantities are nondimensional ($x$ relative to $n_{\mathrm{H}}$
and time in units of $\tau$). Panel $(a)$ shows how the reservoir
of carbon oscillates from $\mathrm{C}$ to $\mathrm{CO}$ and back.
Oscillations are simple and in perfect phase opposition. We note
that the C/CO ratio is of the order of 1, which corresponds to unusual
but observationally found situations with a significant neutral carbon
abundance such as TMC1 (\citealt{1995A&A...294L..17S}). The oscillatory
behavior has indeed been found for a C/O elemental ratio of 1 corresponding
to carbon-rich environments. We also point out that the electronic
fraction displayed in panel $(b)$ of Fig.~\ref{fig:Typical-oscilations.}
oscillates around $4\,10^{-7}$, with an amplitude of $\sim2\,10^{-7}$,
corresponding to a relatively high value for dense interstellar cloud
conditions. The chemical properties of the corresponding solutions
are similar to those of the HIP phase (\citealt{1993ApJ...416L..87L}).
Panel $(c)$ shows how minor species react to this balance. $\mathrm{NH}_{3}$
oscillation amplitudes are mild, but the full isotopic substituted
$\mathrm{ND_{3}}$ shows spectacular amplitude variations, which is
the case for all fully deuterated substituted species. The maximum variation
is over three orders of magnitude for $\mathrm{CD}_{5}^{+}$. Panel $(d)$
shows, for $\mathrm{SH}$, that more complex oscillations are  achievable
that probably emerge from intricate coupling terms in the chemical evolution. 

\begin{figure}
\centering\includegraphics[width=1\columnwidth]{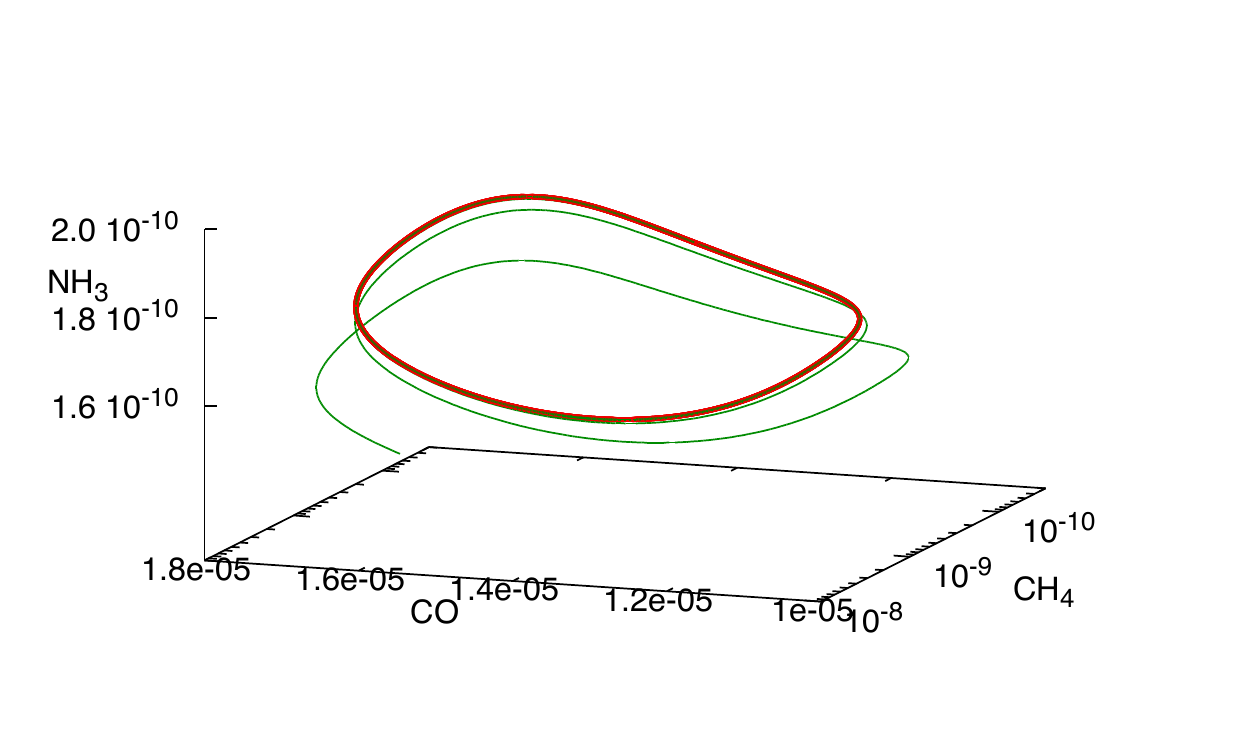}
\caption{Projection of a limit cycle on a 3D slice of phase
space for the fiducial model. Any triple combination of species could
be used in place of $\mathrm{NH}_{3}$, $\mathrm{CH}_{4}$, and $\mathrm{CO}$.\label{fig:Limit-cycle-in}}
\end{figure}

We can also display the dependence between different evolving variables
in phase space. After dissipation of the transitory regime, the phase
trajectory converges towards a limit cycle as displayed for $\mathrm{NH}_{3}$,
$\mathrm{CH}_{4}$, and $\mathrm{CO}$ in Fig.~\ref{fig:Limit-cycle-in}.

\subsection{Initial conditions effects\label{subsec:Initial-conditions-effects}}

\begin{figure}
\centering\includegraphics[width=1\columnwidth]{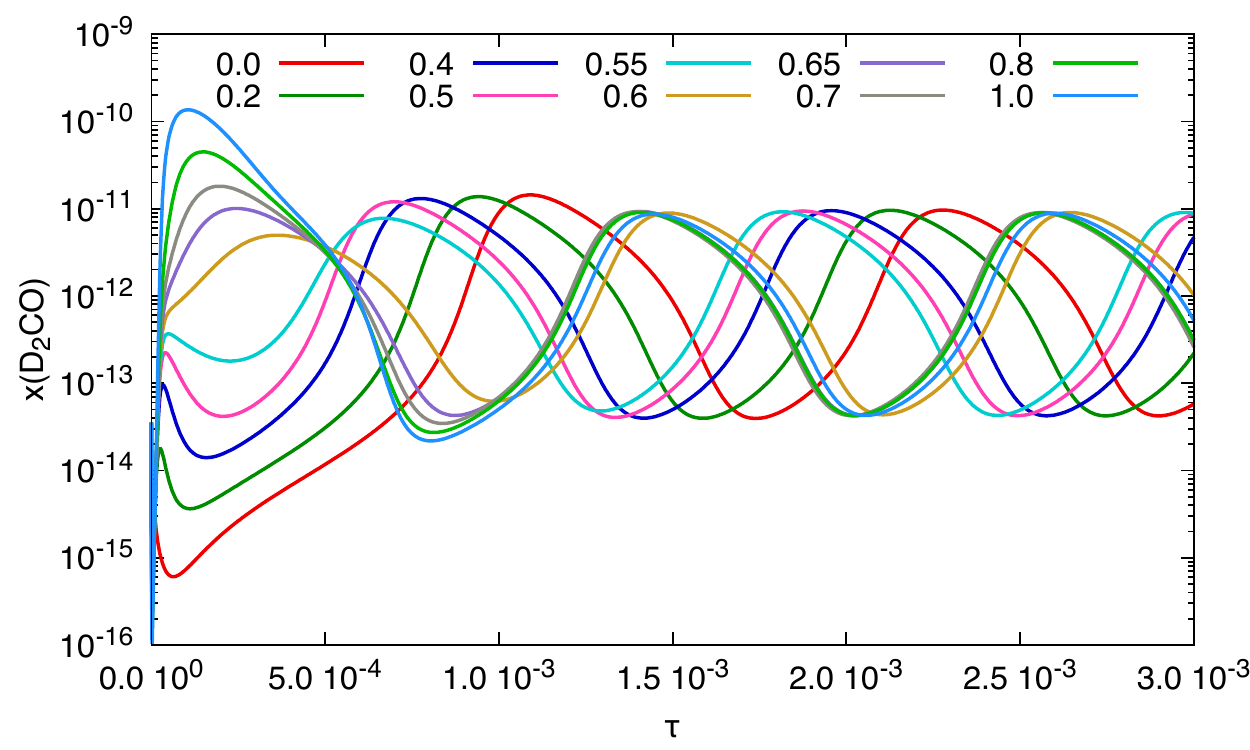}

\caption{Time dependence of D$_{2}$CO for different initial conditions defined
by $\alpha$ (see text) reported on the top with different colors for
the fiducial model. $\alpha=1$ corresponds to all elemental carbon
put on $\mathrm{C}^{+}$ initially.\label{fig:Various-initial-conditions.}}
\end{figure}

Most if not all time-dependent studies use initial conditions
where carbon is totally ionized as $\mathrm{C}^{+}$, sulfur as $\mathrm{S}^{+}$,
whereas oxygen and nitrogen are atomic. Hydrogen is most often taken
as fully molecular but atomic initial conditions are also used sometimes.
These assumptions are obviously far from reality as a dense cloud
results from the previous evolution of diffuse clouds which are found
to be chemically diverse from absorption observations in front of
randomly available continuum radio sources; see for example \citet{2019A&A...627A..95L}.
The partitioning of elemental carbon amongst the different potential
reservoirs, $\mathrm{C}^{+}$, $\mathrm{C}$, and $\mathrm{CO}$ is
much more likely to represent plausible initial conditions. If some
amount of $\mathrm{CO}$ is present as an initial condition, this
impacts the sharing of oxygen as well in order to maintain the full
elemental abundances. Figure~\ref{fig:Various-initial-conditions.}
demonstrates this effect on the time evolution of the fractional abundance of a typical complex
molecule, $\mathrm{D_{2}CO}$\footnote{This species was chosen in order to exhibit a significant amplitude.}.
All initial abundances are identical except for $\mathrm{C}^{+}$,
$\mathrm{O}$, and $\mathrm{CO}$. Initial $\mathrm{C}$ is kept to
a vanishingly low value here, but qualitatively identical results
are obtained if we use $\mathrm{C}$ in place of $\mathrm{C}^{+}$. For
these, the total amount of carbon and oxygen is distributed with a
fraction $\alpha$ on $\mathrm{C}^{+}$ and $\mathrm{O}$ and a fraction
$1-\alpha$ on $\mathrm{CO}$ (recall that $\mathrm{C/O}=1$ here).
We see that $\mathrm{D}_{2}\mathrm{CO}$ oscillations are exactly
the same after about two periods, exhibiting only a phase shift. The
transitory regime occurring before the oscillations take place shows
amplified response to initial conditions that are far from the asymptotic
behavior. But that initial variation can either be towards a stronger
peak (the so-called early-time regime) or a very low abundance,
with all intermediate values possible. We see that the resulting
abundances are entirely a consequence of the choice of initial conditions.
It is therefore very important to select physically plausible conditions,
which excludes the usual choice of a single species found in theoretical
papers. A reasonable solution is for example to run a first case until
a permanent regime is reached, and then to change the control parameters
(e.g., the density) and start from the solution of the previous run.

\subsection{Timescales}

The period of the oscillations can be deduced from the position of
successive maxima. We find a period of $\tau_{0}=1.1562\,10^{-3}$
with $I_{ep}=4$, which translates to $0.73274\,\mathrm{Myr}$ for
$\zeta=5\,10^{-17}\,\mathrm{s}^{-1}$.

To convert from a nondimensional period $\tau_{0}$ to ``real''
time, one needs only to divide by the value of $\zeta$ in $\mathrm{s}^{-1}$.
If $I_{ep}$ is given, then:
\begin{equation}
\frac{\zeta}{10^{-16}\,\mathrm{s}^{-1}} = \frac{n_{\mathrm{H}}}{10^{4}\,\mathrm{cm}^{-3}}\, \frac{1}{I_{ep}}\,,
\end{equation}
and so
\begin{equation}
t_{osc}=\frac{\tau_{0}}{\zeta}=\tau_{0}\,\frac{10^{4}\,\mathrm{cm}^{-3}}{n_{\mathrm{H}}}\,317.1\,I_{ep}\,\mathrm{Myrs}\,.
\end{equation}

\begin{figure}

\centering\includegraphics[width=1\columnwidth]{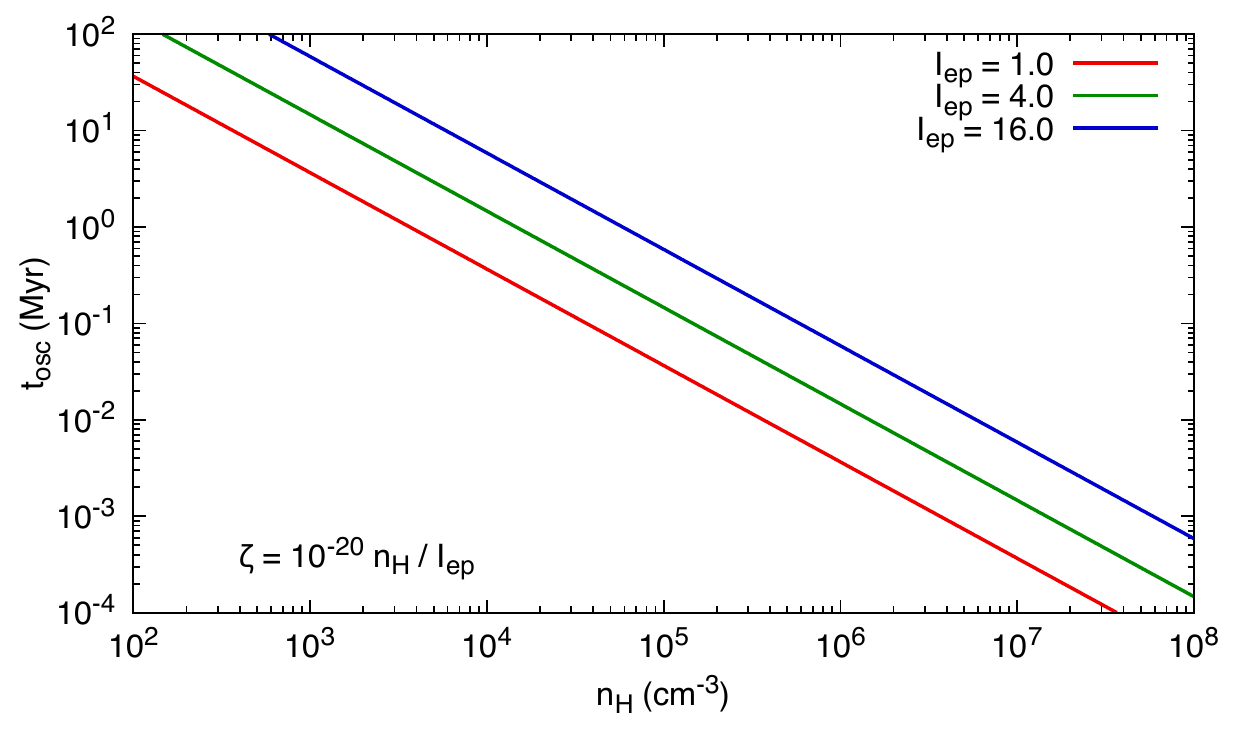}

\caption{Oscillation period as a function of density $n_{\mathrm{H}}$ for
various $I_{ep}$ values.\label{fig:period-as-a}}

\end{figure}

For a given value of $I_{ep}$, the oscillation period decreases as
$1/n_{\mathrm{H}}$. Figure~\ref{fig:period-as-a} shows the relation
between $n_{\mathrm{H}}$ and $t_{osc}$ for the range of $I_{ep}$
found in Sect.~\ref{subsec:Effect-of-Ionis_param}.

\begin{figure}

\centering\includegraphics[width=1\columnwidth]{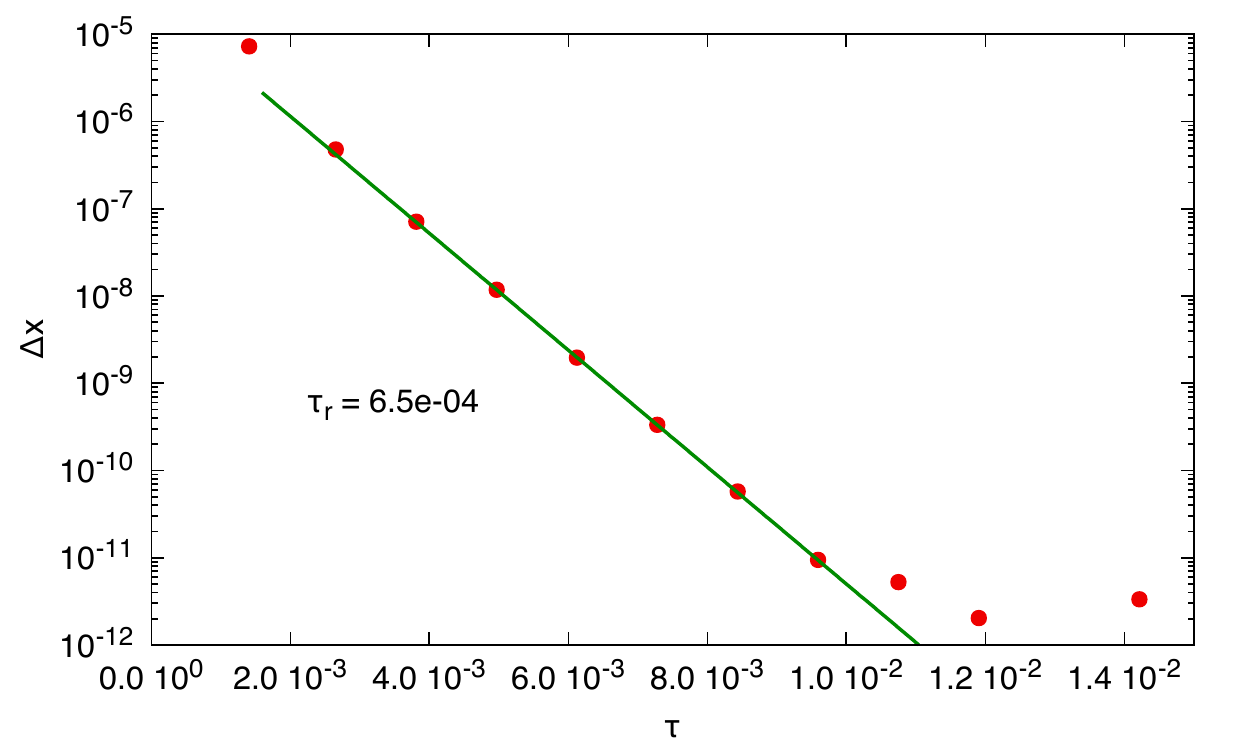}

\caption{Relaxation timescale and period evaluation from successive maxima for the fiducial model.\label{fig:Relaxation-time-scale}}

\end{figure}

Using initial conditions leading to a high first maximum of the time
evolution (e.g., $\alpha=1$ in Fig.~\ref{fig:Various-initial-conditions.}),
we can also study how these maxima converge towards a steady-state
behavior by plotting the differences between successive maxima $\Delta x$
and the asymptotic value. Figure~\ref{fig:Relaxation-time-scale} shows
a typical example \footnote{Here, the fractional abundance of atomic hydrogen, $\mathrm{H}$,
but the specific choice is irrelevant.}. A clear exponential decrease emerges from which we deduce a characteristic
damping time for the transitory regime, $\tau_{r}\simeq6.5\,10^{-4}$,
slightly above a half period.

\begin{figure}
\centering\includegraphics[width=1\columnwidth]{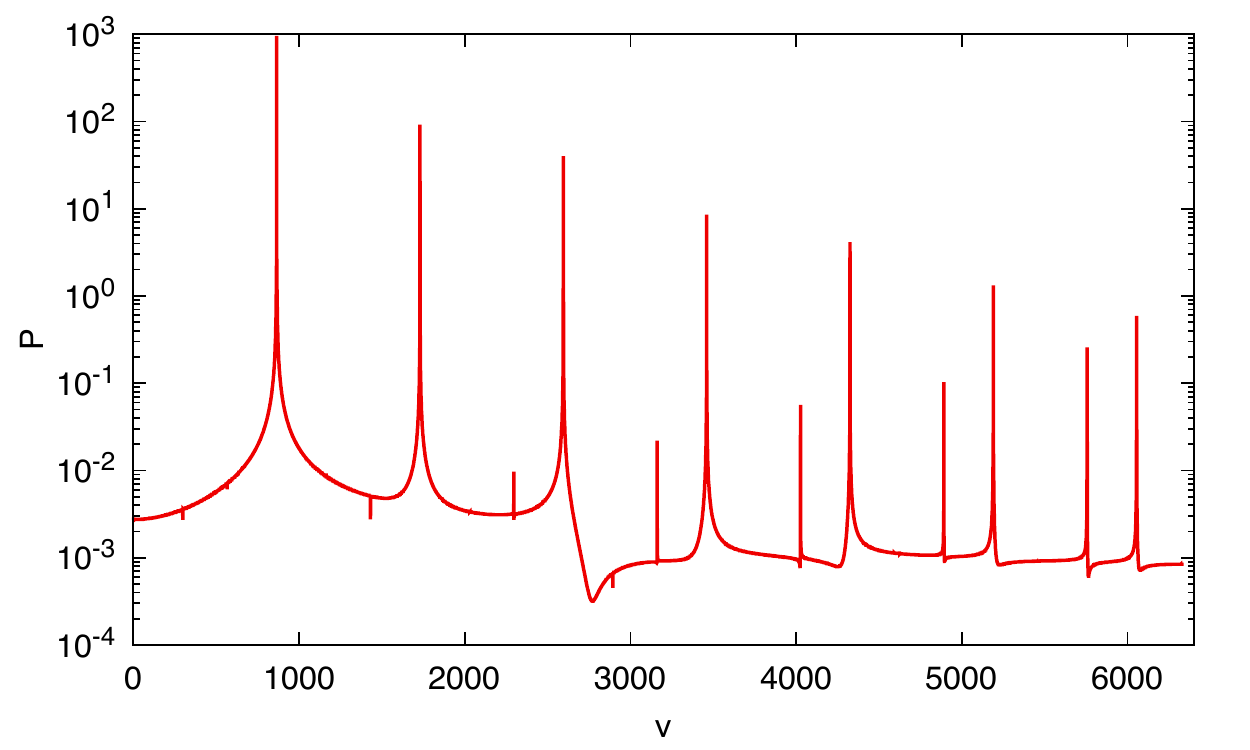}

\includegraphics[width=1\columnwidth]{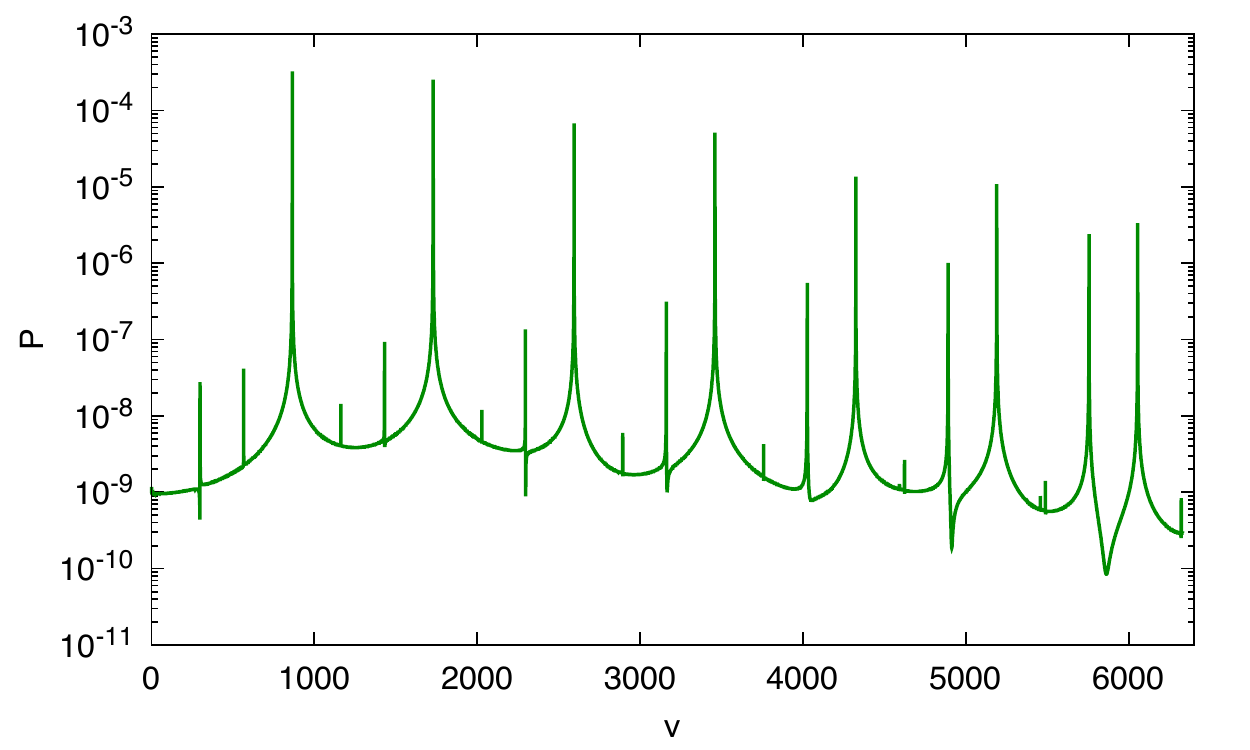}

\caption{Fourier transform power spectrum $P$ of oscillating
fractional abundances in the fiducial model as a function of the dimensionless frequency
parameter $\nu$. Upper panel: $\mathrm{C}$, lower panel $\mathrm{SH}$.\label{fig:Fourier-transform-of}}
\end{figure}

The oscillations are also clearly visible when performing a Fourier
transform of the abundances. The equations were integrated over 1400
periods, which allows us to reach a resolution of $\delta\nu=0.7736$ (unit-less).
The corresponding spectra are shown in Fig.~\ref{fig:Fourier-transform-of},
where the Fourier transform power $P$ of $x\left(\mathrm{C}\right)$
and $x\left(\mathrm{SH}\right)$ evolutions are displayed for comparison.

Although very close, the peak intensities are slightly different at
low frequency. First, the main peak at $\nu_{0}=864.93$ confirms
the value of the period of $\tau_{0}=1.1562\,10^{-3}$ reported above.
All the harmonics at $n\times\nu_{0}$ are very strong. Additional
weaker peaks arise at two other frequencies, $\nu_{1}=298.63$ and
$\nu_{2}=566.27$. However, subsequent peaks at $\nu_{2}+n\times\nu_{0}$
get increasingly stronger when those at $\nu_{1}+n\times\nu_{0}$
disappear. This is particularly perceptible for $\mathrm{SH}$. We
also point out that $\nu_{1}$ is close to $\nu_{0}/3$ and $\nu_{2}$
is close to $2\nu_{0}/3$ but the difference between these fractions
is markedly greater than the frequency resolution, and is therefore significant.
These additional peaks are therefore not sub-harmonics of the main frequency.

\section{Dependence of the oscillatory behavior on the control parameters\label{sec:Variation-of-control}}

Such a chemical evolution could affect the dynamics of the environment
if its occurrence complies with likely observable conditions, as discussed in Sect.~\ref{sec:Discussion}.
Here, we study within which range of control parameters oscillations are obtained.

\subsection{Effect of the ionization efficiency parameter $I_{ep}$\label{subsec:Effect-of-Ionis_param}}

The standard method to find the extent of an oscillatory zone in parameter
space is the following:
\begin{enumerate}
\item Fix a value of the parameter $I_{ep}$ and find the exact value of
$T$ at the bifurcation. This should be achieved by computing the
eigenvalues of the Jacobian matrix and solving for $\lambda_{m}\left(T\right)=0$,
where $\lambda_{m}$ is the highest eigenvalue real part.
\item Then use a continuation scheme to follow the bifurcation point in
parameter space. To this end, we require that a variation $\delta I_{ep}$
of $I_{ep}$ leads to a variation $\delta T$ of $T$ to still be
at a bifurcation point.
\begin{equation}
\lambda_{m}\left(I_{ep},T\right)=0\,\Rightarrow\,\lambda_{m}\left(I_{ep}+\delta I_{ep},T+\delta T\right)=0\,.
\end{equation}
\end{enumerate}
Then, to first order:
\begin{equation}
\lambda_{m}\left(I_{ep}+\delta I_{ep},T+\delta T\right)\simeq\lambda_{m}\left(I_{ep},T\right)+\frac{\partial\lambda_{m}}{\partial I_{ep}}\,\delta I_{ep}+\frac{\partial\lambda_{m}}{\partial T}\,\delta T\,.
\end{equation}

If ${\displaystyle \frac{\partial\lambda_{m}}{\partial T}}\neq0$,
we can solve the 1D ordinary differential equation (EDO):
\begin{equation}
\frac{dT}{dI_{ep}}=-\frac{\partial\lambda_{m}}{\partial I_{ep}}/\frac{\partial\lambda_{m}}{\partial T}\,.
\end{equation}

Unfortunately, this scheme breaks down in the present case. The Jacobian
matrix happens to be very ill-conditioned, and the computation of
the eigenvalues becomes unstable close to the bifurcation. This has
been confirmed using two different numerical routine libraries:
the latest LAPACK version of DGEEVX, and the oldest but quadruple
precision version of EISPACK routine ``rg''. Both routines agree
very well on all eigenvalues, except in the vicinity of a bifurcation
point where the smallest eigenvalues in norm are unreliable\footnote{We note, as a curiosity, that the sum of the times deduced from the
imaginary parts of the eigenvalues is equal to the period of oscillations
within the numerical uncertainty. Formally, there is no clear reason
why the correct linear combination should be the one with all coefficients
equal to $1$.}.

\begin{figure}
\centering\includegraphics[width=1\columnwidth]{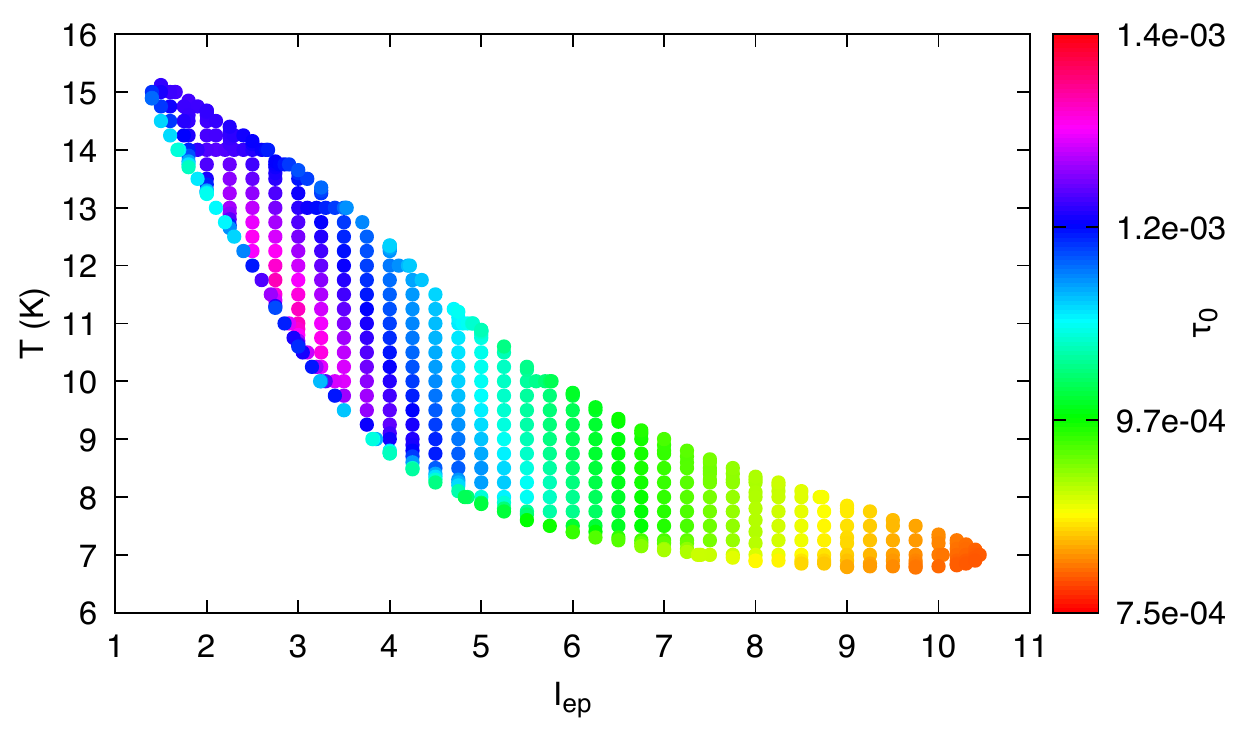}

\caption{Oscillation domain in the $I_{ep}-T$ plane. The
oscillation period is color-coded.\label{fig:Oscillation-period-in}}
\end{figure}

We therefore resorted to a less accurate but robust method to explore parameter
space: brute force. Figure~\ref{fig:Oscillation-period-in} shows the
domain of oscillations in the $I_{p}-T$ parameter space obtained
from systematic sampling, together with the oscillation period represented by
the color coding. The domain of oscillation spans a narrow
range of temperatures between about $7$ and $15\,\mathrm{K}$ with
ionization efficiency parameters between $1.3$ and $10.4$. For a
typical cosmic-ray ionization parameter $\zeta=5\,10^{-17}\,\mathrm{s}^{-1}$,
the range of density is $5\,10^{3}-5\,10^{4}\,\mathrm{cm}^{-3}$.
We stress that all these values are compatible with physical conditions
found in interstellar clouds. The oscillatory period varies by less
than a factor of two, between $450$ and $900\,\mathrm{kyr}$
for this value of cosmic-ray ionization rate. Such values are
also similar to the expected evolution time of interstellar clouds.
Small ionization efficiency parameters lead to larger values of the
oscillation period. We can therefore consider that these findings are much
more than a numerical curiosity and could significantly impact our
knowledge of the interstellar medium.

\subsection{Effect of the temperature\label{subsec:Effect-of-temperature}}

\begin{figure}
\centering\includegraphics[width=1\columnwidth]{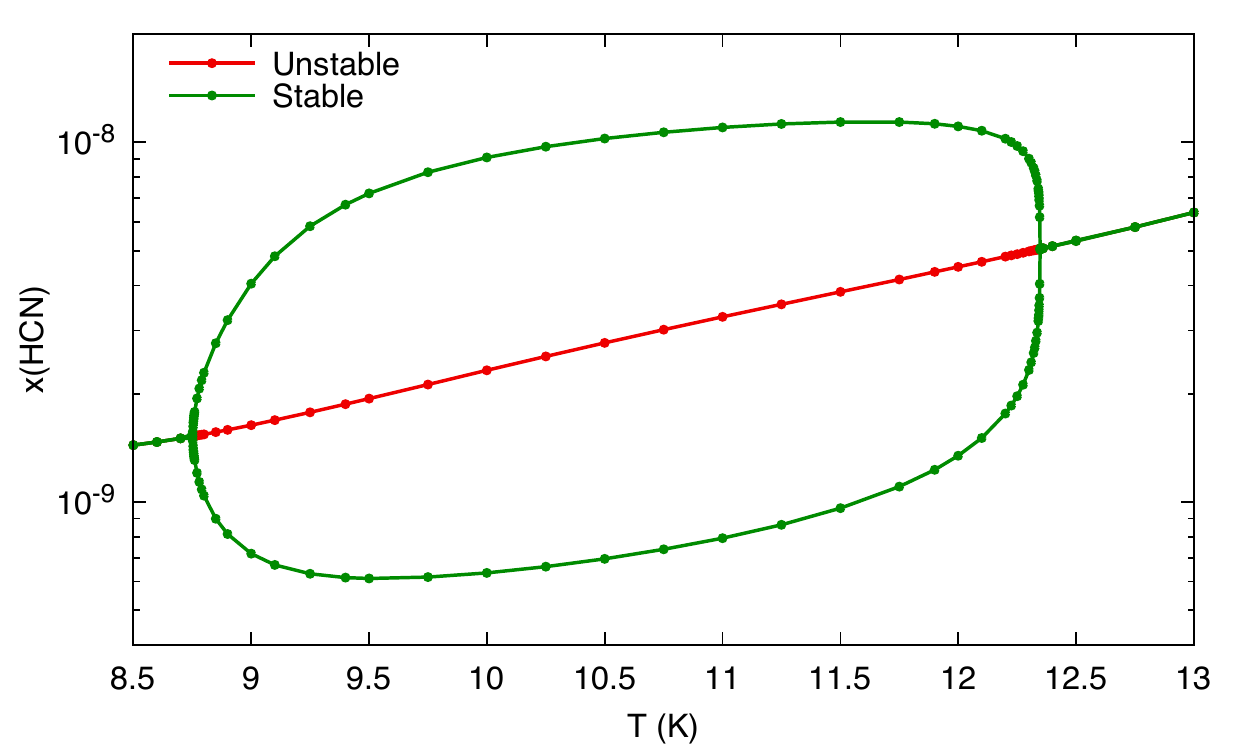}

\caption{Bifurcation diagram for $\mathrm{HCN}$.{} Within
the unstable fixed point region, the two outer curves show the range
of oscillations (minimum and maximum values).\label{fig:Bifurcation-diagram-for}}
\end{figure}

Keeping other parameters from Table~\ref{tab:Reference-model-parameters.}
fixed, we now explore the range of temperatures over which oscillations
are found. Figure~\ref{fig:Bifurcation-diagram-for} is a bifurcation
diagram displaying the steady-state solutions when $T$ is varied
from $8.5\,\mathrm{K}$ to $13\,\mathrm{K}$, keeping $I_{ep}=4$.
A single fixed point is obtained for $T_{c1}<8.752\,\mathrm{K}$ and
$T_{c2}>12.347\,\mathrm{K}$, which is exhibited in green. Stable
oscillations are present in between and the corresponding maxima and
minima values of the fractional abundances are also reported in green.
The fixed point itself does not disappear, but becomes unstable, and
this is displayed in red in Fig.~\ref{fig:Bifurcation-diagram-for}.
This behavior is standard, and representative of a Hopf bifurcation,
as detailed for example in \citet{gray1994chemical}.

\begin{figure}
\centering\includegraphics[width=1\columnwidth]{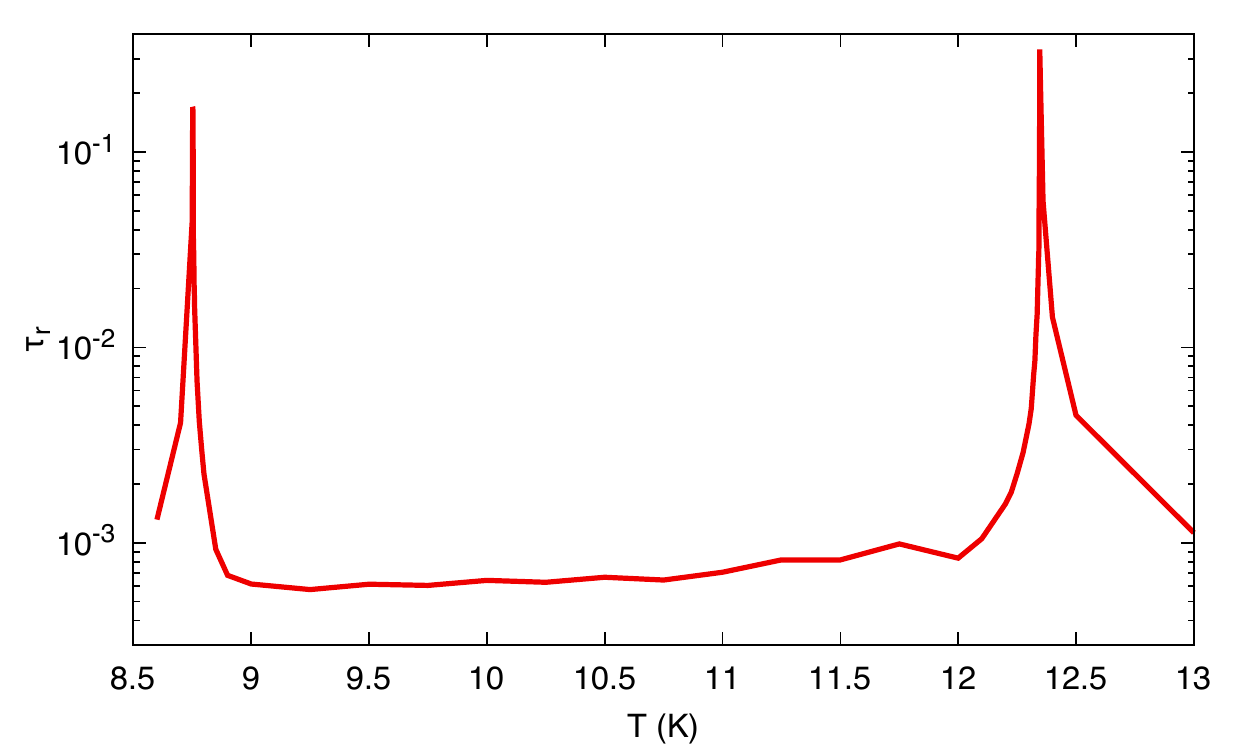}

\caption{Relaxation time of initial conditions. We highlight the large uncertainties
far from the bifurcation points. Conversely, close
to $T_{c1}$ and $T_{c2}$ the large relaxation time allows for a
very accurate determination.\label{fig:Damping-time-of}}
\end{figure}

This effect of temperature
is illustrated using $\mathrm{HCN}$, but the behavior is generic
and identical for all species. Below $T_{c1}=8.752\,\mathrm{K}$, there
is a stable fixed point. Starting from various (but not necessarily
any) initial conditions, the system converges towards it. However,
convergence is slower and slower corresponding to an increasing relaxation
time of the initial conditions, as seen in Fig.~\ref{fig:Damping-time-of},
which displays the relaxation time deduced from the decline of successive maxima
as a function of temperature.

\begin{figure}
\centering\includegraphics[width=1\columnwidth]{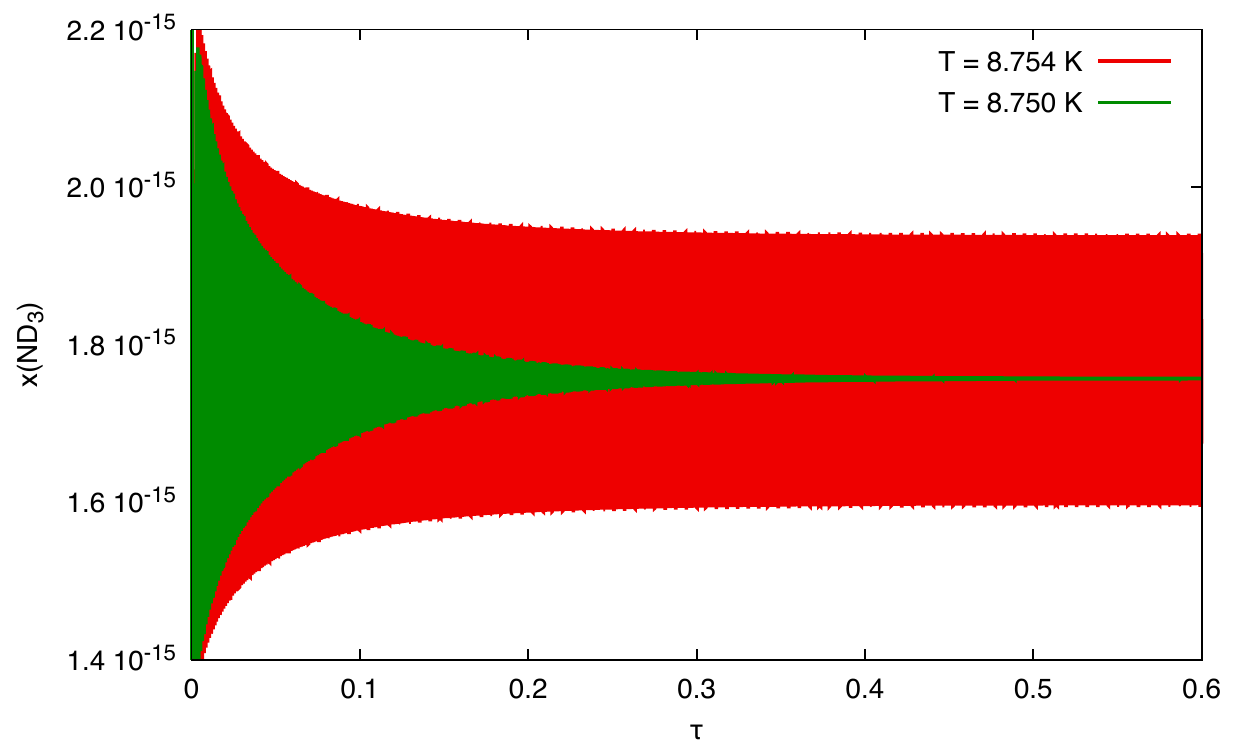}

\caption{$\mathrm{ND}_{3}$ evolution before and after the first bifurcation
point.\label{fig:-evolution-before}}
\end{figure}

This is the critical slow-down phenomenon, characteristic of bifurcations.
Close to $T_{c1}$, the fixed point becomes unstable, and oscillations
appear through a Hopf bifurcation. Their amplitudes are initially
vanishingly small, but they grow fast as $T$ is increased. The time
evolution for two values of the temperature that bracket the bifurcation
point is shown in Fig.~\ref{fig:-evolution-before}. In the close
vicinity of $T_{c1}$ and for $T<T_{c1}$, a nice damped oscillatory
structure is shown for the $\mathrm{ND}_{3}$ fractional abundance,
which converges slowly towards the stable fixed point (in green on
Fig.~\ref{fig:-evolution-before}). The red curve displays the sustained
oscillations taking place immediately after $T_{c1}$. One can see
that many oscillations are necessary before reaching the permanent
regime.

The amplitude of the oscillations remains significant until the vicinity
of $T_{c2}=12.347\,\mathrm{K}$ where a second Hopf bifurcation occurs.
Beyond this second bifurcation point, the fixed point is stable again.
A fixed point is still present between the two bifurcation points
but the corresponding solution is unstable. These points are displayed in red
in Fig.~\ref{fig:Bifurcation-diagram-for} and were computed
using a Newton-Raphson scheme by directly solving the $d/dt=0$
coupled chemical equations.

The critical slow-down in the vicinity of a bifurcation point is important,
because it leads to timescales that are totally unrelated to any
other characteristic times of the system, either chemical or dynamical.
Hence, the system may seem to behave strangely given the chosen parameters.
This is simply the direct consequence of a close-by bifurcation point
in parameter space. It is also remarkable that the range of temperatures
where the oscillatory behavior takes place is within the values derived
from interstellar cloud observations.

The value $T_{c2}$ is driven by our hypothesis on $(O/P)$, the $\mathrm{H}_{2}$
ortho-to-para ratio. Here, we assume that it is the thermal
equilibrium value, where $T$ is the gas temperature. Increasing $T$
increases $(O/P)$, which has a dramatic effect on the
$\mathrm{N}^{+}+\mathrm{H}_{2}\rightarrow\mathrm{NH}^{+}+\mathrm{H}$
reaction rate coefficient (see discussion on the chemistry in Sect.~\ref{subsec:Chemistry}
and Appendix~\ref{sec:Nitrogen-chemistry} and on the $(O/P)$ value
in Sect.~\ref{subsec:OsP_ratio}).

\subsection{Effect of the elemental abundances}

Elemental abundances in the gas phase constitute another set of control
parameters. We focus our study on the most abundant elements after hydrogen and helium,
namely $\mathrm{C}$, $\mathrm{N}$, and $\mathrm{O}$, and introduce
a relation between $\mathrm{\delta}_{\mathrm{C}}$ and $\mathrm{\delta}_{\mathrm{O}}$.
Instead of varying the latter independently, we explore the influence
of the $\mathrm{\delta}_{\mathrm{C}}/\mathrm{\delta}_{\mathrm{O}}$
ratio at constant total $\mathrm{\delta}_{C}+\mathrm{\delta}_{\mathrm{O}}$
(total metallicity) or the influence of metallicity for a constant
$\mathrm{\delta}_{\mathrm{C}}/\mathrm{\delta}_{\mathrm{O}}=1$.

\begin{figure}
\centering\includegraphics[width=1\columnwidth]{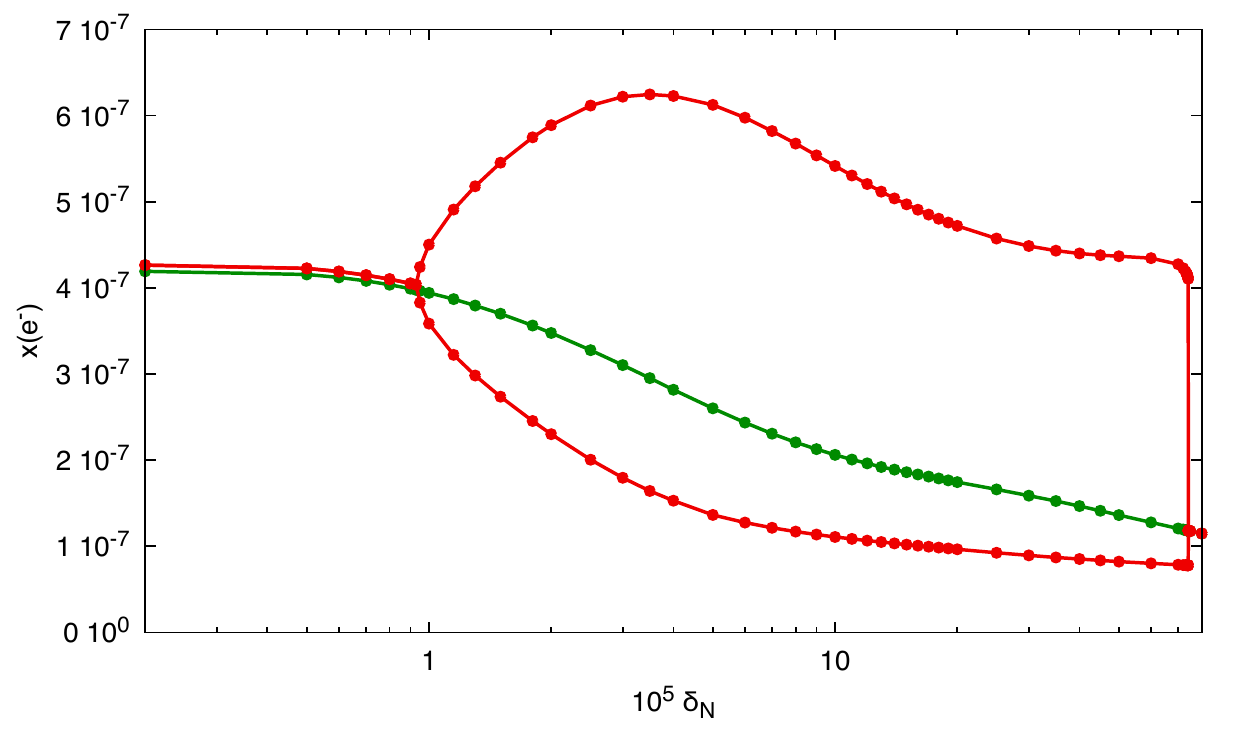}
\caption{Bifurcation diagram for $\mathrm{\delta}_{\mathrm{N}}$.
Note a small error in the value of the fixed point computed with the
Newton-Raphson scheme below the lowest Hopf bifurcation.\label{fig:Influence-of-delta_N}}
\end{figure}

Figure~\ref{fig:Influence-of-delta_N} shows a bifurcation diagram
for $\mathrm{\delta}_{\mathrm{N}}$. All other parameters are kept
at the values of our fiducial model (cf Table~\ref{tab:Reference-model-parameters.}).
As the density $n_{\mathrm{H}}$ is kept constant, we chose the
ionization degree of the gas as a representative variable, as it does
not require a rescaling by elemental abundances. Two Hopf bifurcations
are present, with oscillations maintained up to a very high (and somewhat unrealistic)
abundance value for nitrogen (over $7\,10^{-4}$).

\begin{figure}
\centering\includegraphics[width=1\columnwidth]{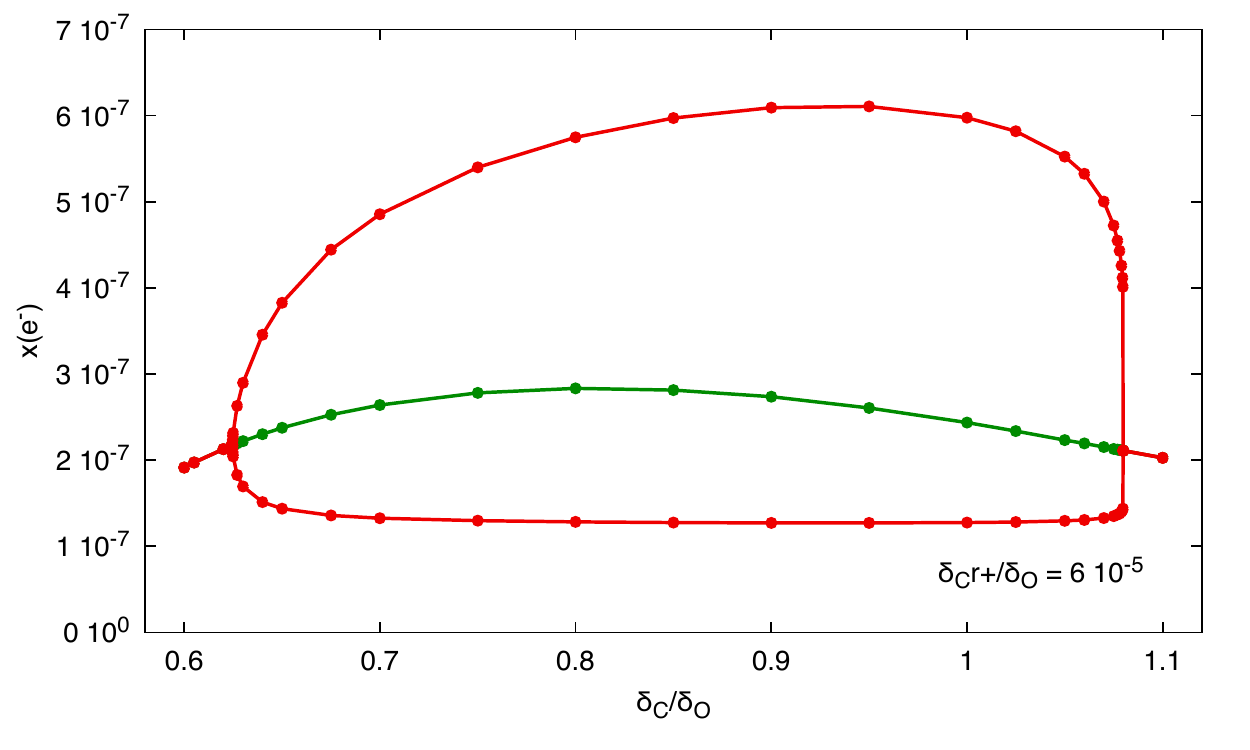}
\caption{Bifurcation diagram for $\mathrm{\delta}_{\mathrm{C}}/\mathrm{\delta}_{\mathrm{O}}$.\label{fig:Influence-of-CsO}}
\end{figure}

\begin{figure}
\centering\includegraphics[width=1\columnwidth]{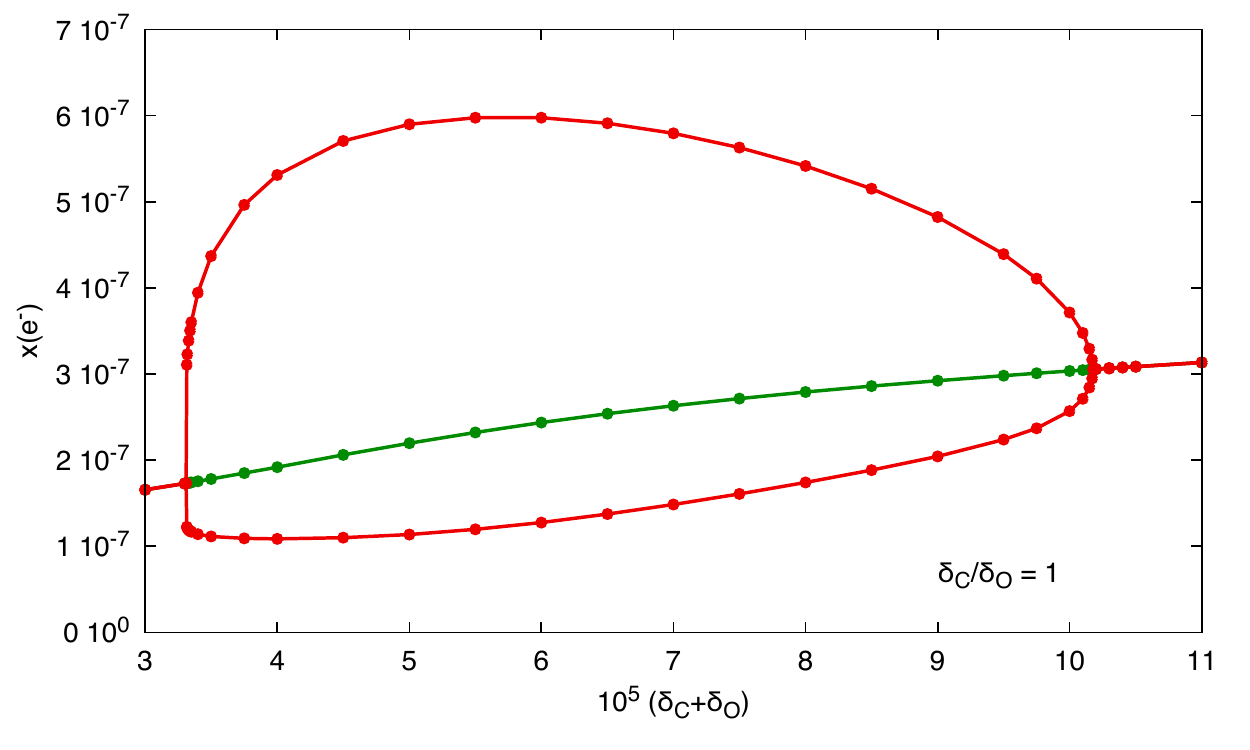}
\caption{Bifurcation diagram for the total metallicity with
varying $\delta_{\mathrm{C}}+\delta_{\mathrm{O}}$. Other abundances
are kept as in Table~\ref{tab:Reference-model-parameters.}.\label{fig:Influence-of-tOpC}}
\end{figure}

Keeping $\mathrm{\delta}_{\mathrm{N}}$ at its fiducial value ($6\,10^{-5}$),
Fig.~\ref{fig:Influence-of-CsO} shows that oscillations encompass
typical values of $\mathrm{\delta}_{\mathrm{C}}/\mathrm{\delta}_{\mathrm{O}}$
and not only the selected value of $1$ we used up to here.
Figure~\ref{fig:Influence-of-tOpC}
shows in a similar way that the total metallicity can be varied up
to values found in many environments, although somewhat lower than
in the solar vicinity.

We conclude that the occurrence of oscillations does not require very
specific values of the elemental abundances, although our studies
are more representative of low-metallicity or depleted environments.

The chance occurrence of a spectacular critical slow-down found during
the exploration of this parameter space is presented in Appendix~\ref{sec:Spectacular-critical-slow}.

\section{Discussion\label{sec:Discussion}}

\subsection{Chemistry\label{subsec:Chemistry}}

\begin{figure}
\centering\includegraphics[width=1\columnwidth]{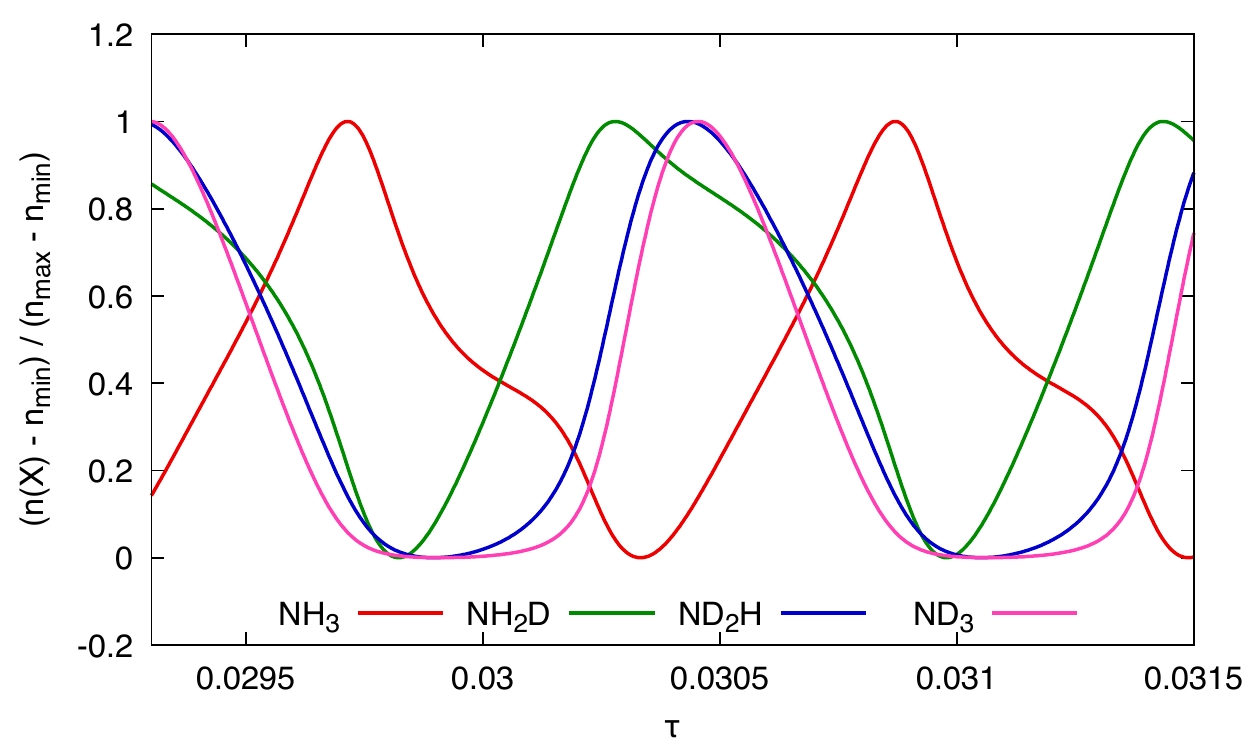}

\caption{Scaled oscillations of $\mathrm{NH}_{3}$ and its deuterated isotopologs.
The amplitude of the oscillations is fixed to $[0:1]$ for
each species as described in the text.\label{fig:Scaled-oscillations-of}}

\end{figure}

Elucidating the chemical mechanisms responsible for oscillations is
much more challenging than for steady-state results. There is indeed
no set of equations to solve that would provide the answer. One has
to follow the time variations of leading species to point to the relevant
interactions. Here, we do not know which are the leading species.
We can however derive some chemical properties that comply with the
oscillatory behavior:
\begin{enumerate}
\item The full evolution remains within HIP chemistry. This corresponds
to a significant electronic fraction ($\sim$ a few $10^{-7}$) which
further implies the presence of $\mathrm{C}$ at a level close to
that of $\mathrm{CO}$.
\item Full deuteration appears compelling (within our set of chemical reactions).
This implies that long chains of reactions exist leading from some
very simple species (say, e.g., $\mathrm{N}^{+}$) to fully deuterated
species (say, e.g., $\mathrm{ND}_{3}$, $\mathrm{ND}_{4}^{+}$). It
may be possible that other long chains of reactions also lead to oscillations,
for example the formation of cyanopolyines, but we did not explore this point
further.
\item Nitrogen chemistry plays a central role. The slightly endothermic
$\mathrm{N}^{+}+\mathrm{H}_{2}\rightarrow\mathrm{NH}^{+}+\mathrm{H}$
reaction, which is subject to large uncertainties (\citealt{dislaire:12,1988JChPh..89.2041M}),
has a huge impact on the collective behavior.
\end{enumerate}
A first clue as to the underlying mechanism is seen from the successive
maxima of chemically linked species. Figure~\ref{fig:Scaled-oscillations-of}
shows the sequence of different deuterated compounds of $\mathrm{NH}_{3}$
as a function of time. The abundances are rescaled to the (dimensionless)
interval $[0:1]$ through $\frac{(n(t)-n_{min})}{(n_{max}-n_{min)}}$
to enable comparison. We first observe the peak of $\mathrm{NH}_{3}$
itself. Then, offset by nearly half a period, we find the peak of
$\mathrm{NH_{2}D}$, which is closely followed by that of $\mathrm{ND_{2}H}$,
and finally by that of $\mathrm{ND}_{3}$. Meanwhile, $\mathrm{NH}_{3}$ has
time to decrease to its minimum, which quenches the deuteration process.
The deuterated species are then destroyed more efficiently than they
are formed, and they fall to very low levels. The process then starts
again.

\begin{figure}

\centering\includegraphics[width=1\columnwidth]{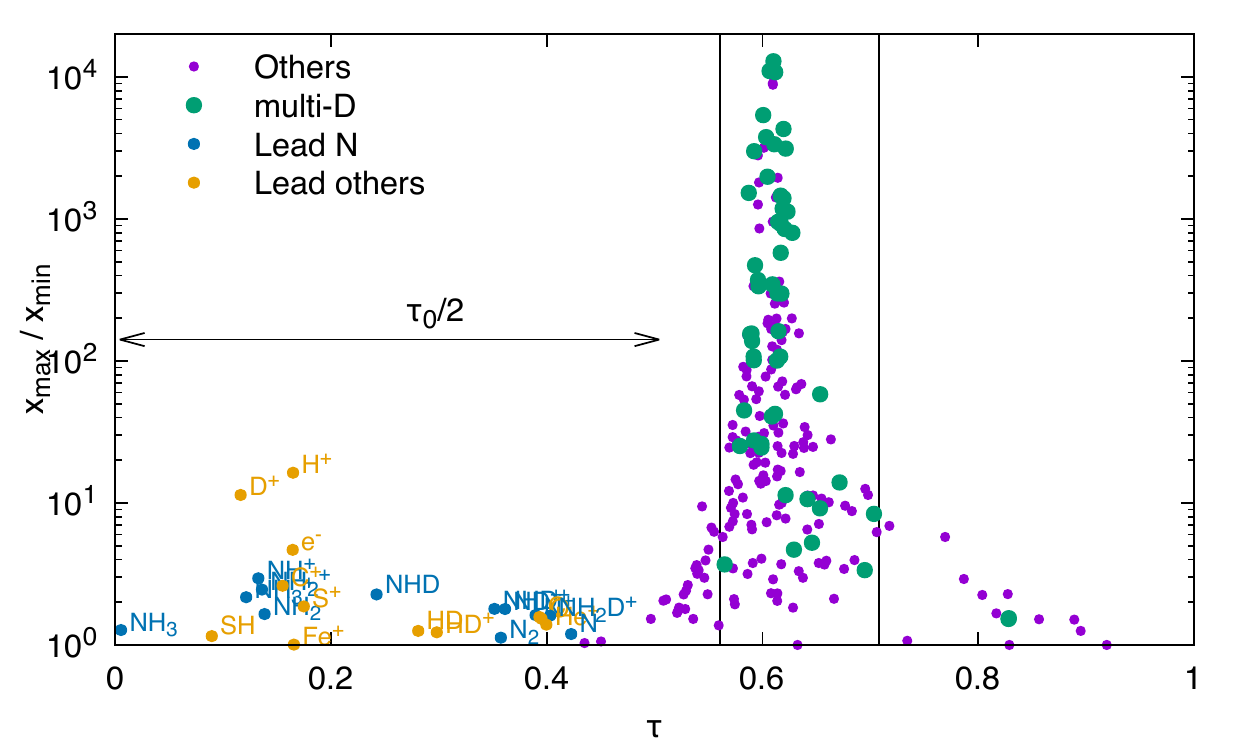}

\caption{Ratio of maximum to minimum abundances as a function of phase shift
with respect to $\mathrm{NH}_{3}$. ``multi-D'' are molecules with two or more
$\mathrm{D}$. The two vertical lines delineate
the time range where multideuterated species peak.\label{fig:Ratio-of-maximum}}

\end{figure}

Figure~\ref{fig:Ratio-of-maximum} shows the ratio of the maximum
to minimum abundances of all species as a function of their
peak abundance time with respect to that of $\mathrm{NH}_{3}$. The
peak of $\mathrm{NH}_{3}$ is well separated from all other species
(by at least one-tenth of a period), which is a strong indication that
it plays a particular role. This latter peak is followed by a small number of ``lead''
molecules, half of which are hydrogenated nitrogen with no or one
deuterium. Then, at about $0.6\,\tau_{0}$, most species peak, with
an amplitude that may exceed a factor of $10^{4}$. This is where
all multiply deuterated molecules are found, with the single exception
of $\mathrm{D}_{2}^{+}$. This peak extends over a limited time interval,
which expresses the trigger of successive interconnected maxima. We
interpret these features as a signature of the importance of nitrogen
deuteration in the mechanism responsible for oscillations.

\subsubsection{$\mathrm{NH}_{3}$/$\mathrm{CN}$ connection }

\begin{figure*}
\centering\includegraphics[width=1\textwidth]{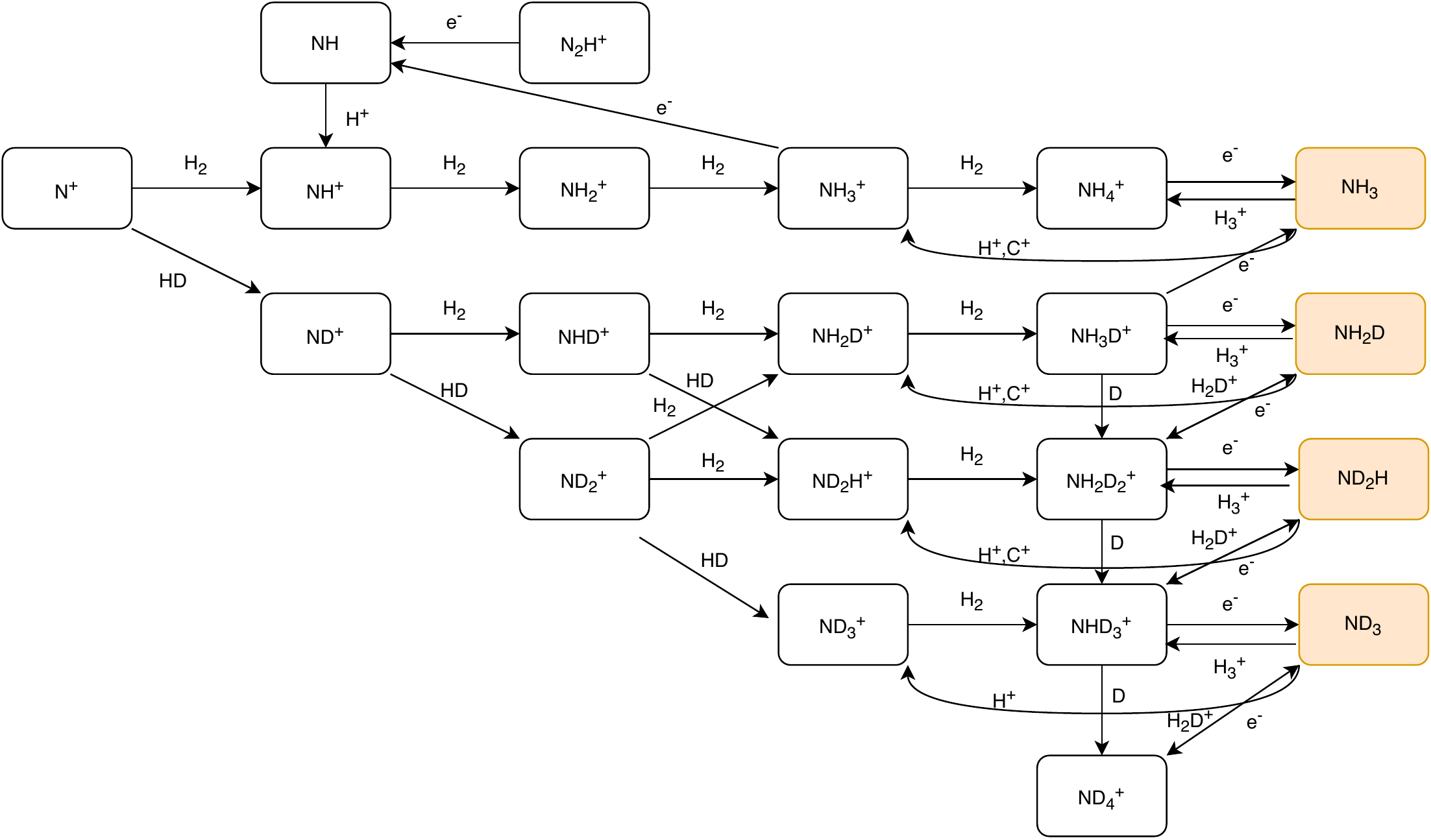}

\caption{$\mathrm{NH}_{3}$ and deuterated isotopolog formation.\label{fig:Set-1:-NH3-formation.}}
\end{figure*}

Formation of $\mathrm{NH}_{3}$ and its deuterated substitutes proceeds
progressively in time as shown in Fig.~\ref{fig:Scaled-oscillations-of}.
The initial step is provided by the reaction of $\mathrm{N}^{+}$
with $\mathrm{H}_{2}$, $\mathrm{HD}$, and $\mathrm{D}_{2}$ as shown
in Fig.~\ref{fig:Set-1:-NH3-formation.}.

\begin{figure*}
\centering\includegraphics[width=1\textwidth]{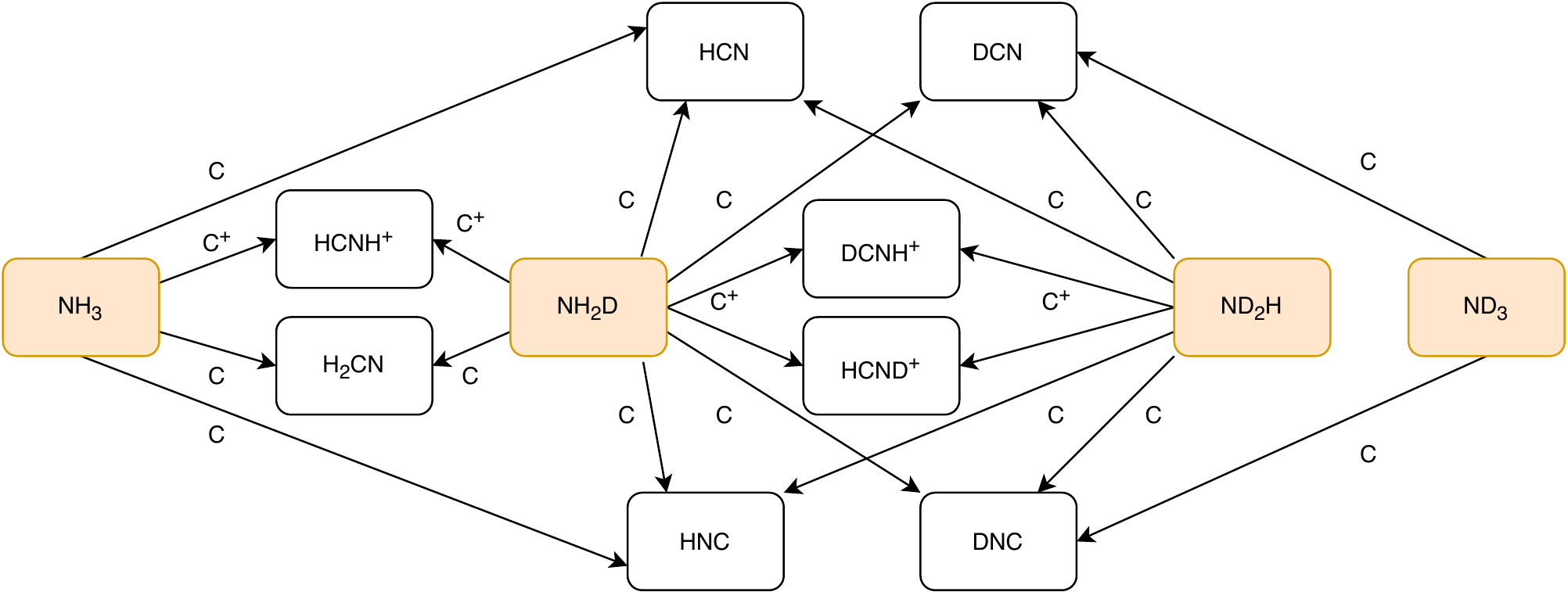}

\caption{$\mathrm{NH}_{3}$ and deuterated isotopolog destruction.
The dissociative recombination reactions of $\mathrm{HCNH}^{+}$ and deuterated
substitutes are not displayed.
\label{fig:Set-4:-NH3-Destruction}}
\end{figure*}

\begin{figure*}
\centering\includegraphics[width=1\textwidth]{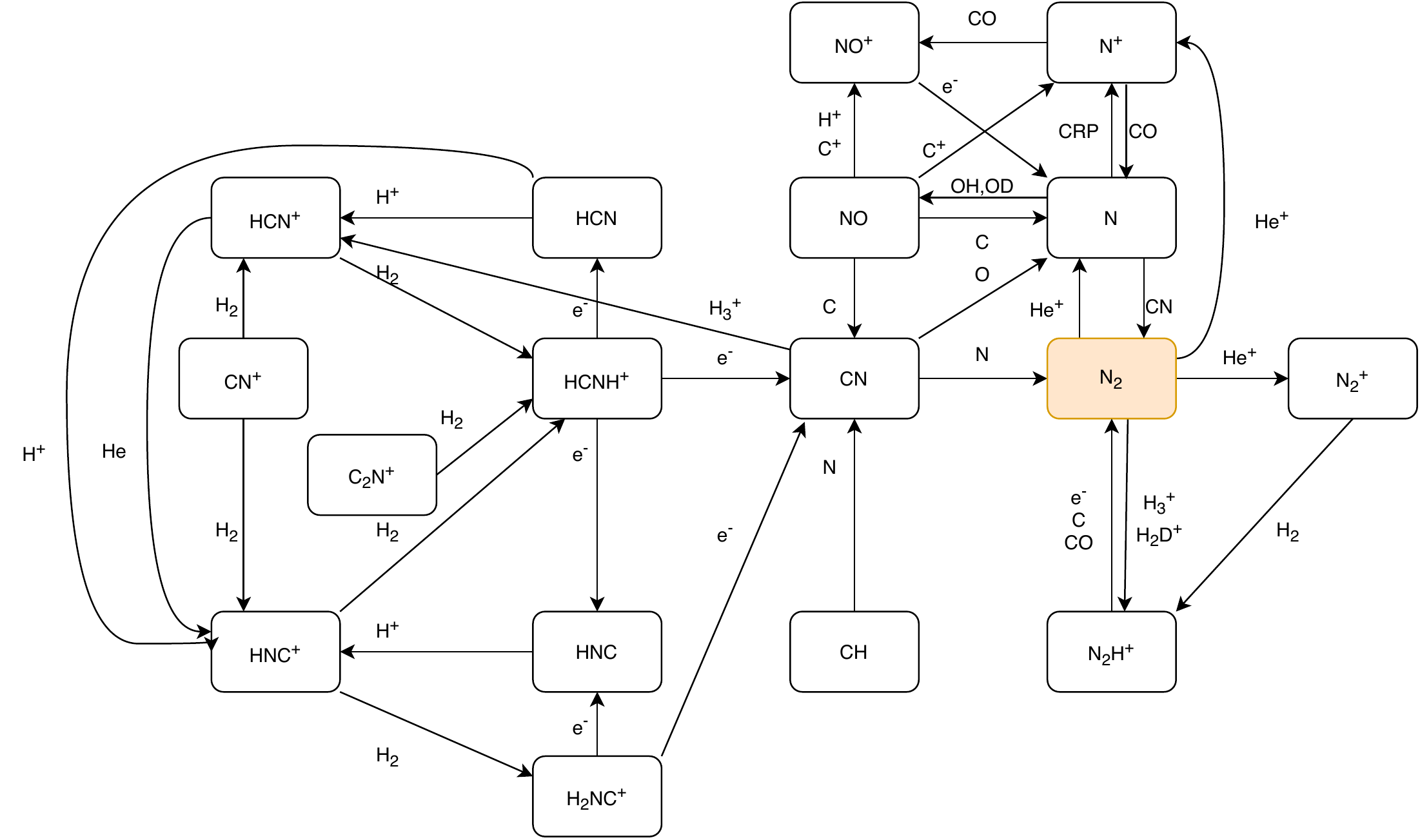}

\caption{$\mathrm{CN}$ links destruction and nitrogen recycling.\label{fig:Set-5:-CN-Recycling}}
\end{figure*}

In the HIP environment considered here, atomic carbon $\mathrm{C}$
(\citealt{2015ApJ...812..106B,2015ApJ...812..107H}) and ionized carbon
$\mathrm{C}^{+}$ are the main destruction agents (see Fig.~\ref{fig:Set-4:-NH3-Destruction})
which finally lead to the formation of a $\mathrm{CN}$ bond within
a large variety of species ($\mathrm{HCN}$, $\mathrm{HNC}$, $\mathrm{HCNH}^{+}$,..),
as displayed in Fig.~\ref{fig:Set-5:-CN-Recycling}.

Focusing on these $\mathrm{H}$, $\mathrm{C}$, and $\mathrm{N}$ compounds
as displayed on the left of Fig.~\ref{fig:Set-5:-CN-Recycling},
we see that the $\mathrm{CN}$ radical emerges. Reactions of
$\mathrm{CN}$ with $\mathrm{C}$ and $\mathrm{O}$ then lead back to atomic
$\mathrm{N}$ whereas the reaction with $\mathrm{N}$ produces $\mathrm{N}_{2}$
as found experimentally by \citet{2012PNAS..10910233D}. $\mathrm{N}^{+}$
is subsequently restored via reactions of $\mathrm{N}_{2}$ with $\mathrm{He}^{+}$ and
cosmic-ray ionization of atomic $\mathrm{N}$, and the full cycle starts
again.

During the whole process, minor reactions may create leaks towards
other branches of the full chemical set (e.g., through $\mathrm{NO}$
for a coupling to oxygen chemistry), and so it is very sensitive, which
explains why we find oscillations only within a restricted part of
parameter space. We further study nitrogen chemistry in Appendix~\ref{sec:Nitrogen-chemistry}.

\subsection{Thermal balance}

Our current model does not include a sophisticated treatment of thermal
balance, as found for example in the Meudon PDR code (\citealt{2006ApJS..164..506L}).
However, simple analytic formulae of the main heating and cooling processes
are included, following \citet{2015A&A...578A..63F}, that allow the
derivation of the temperature within a given equation of state, namely
$n=Cte$ here, with $n$ the density of the gas, that is,
$n=n\left(\mathrm{H}\right)+n\left(\mathrm{H}_{2}\right)+n\left(\mathrm{He}\right)+...$.

Solving for thermal balance removes one control parameter from the
system, as temperature is now computed by solving its evolution
equation and is not set arbitrarily. However, the cooling and heating terms
do not scale in the same way with $\zeta$ and $n_{\mathrm{H}}$,
and so the ionization efficiency $I_{ep}$ cannot be used as a single
parameter. The number of control parameters is therefore unchanged.

\begin{figure}

\centering\includegraphics[width=1\columnwidth]{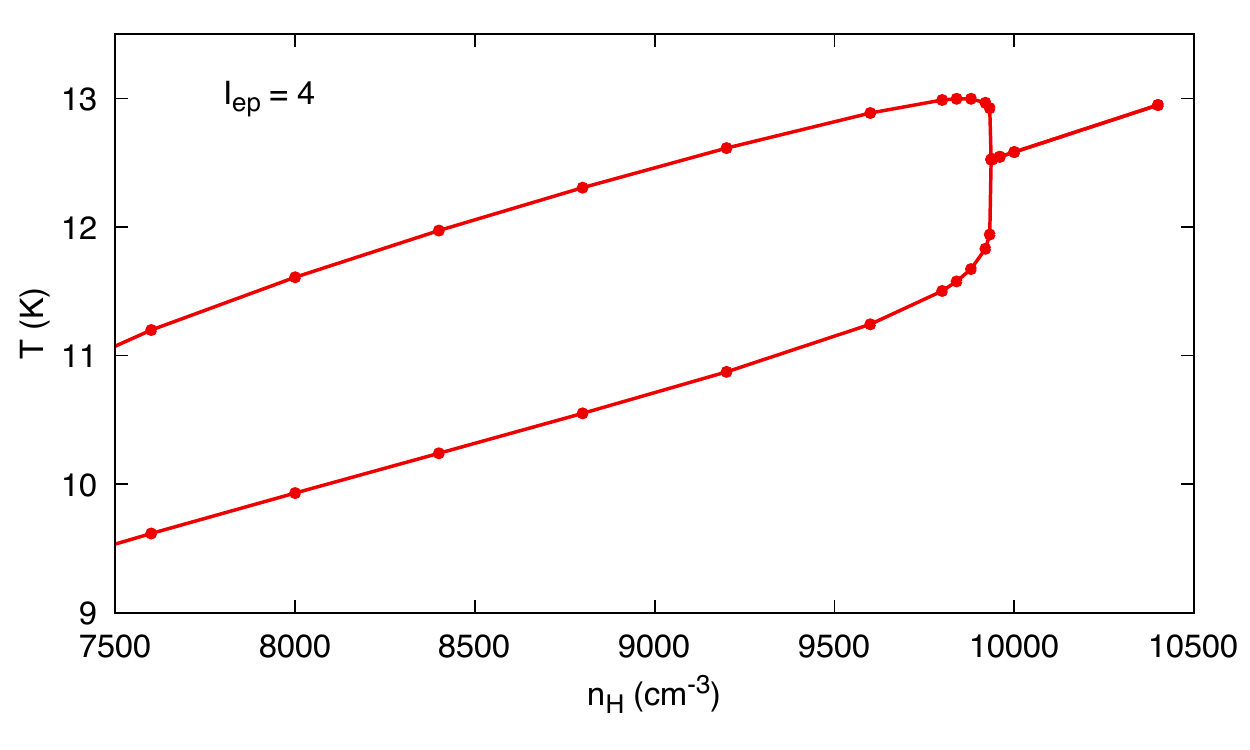}

\caption{Bifurcation diagram for temperatures at
 $I_{ep}=4$.\label{fig:Thermal-oscillations-for}}

\end{figure}

We did not conduct a full exploration of that parameter space, but
we checked that thermal oscillations can be found and are fully compatible
with the mechanism described here. For $I_{ep}=4$, Fig.~\ref{fig:Thermal-oscillations-for}
shows a Hopf bifurcation close to $n_{\mathrm{H}}=9.5\,10^{3}\,\mathrm{cm}^{-3}$.
We kept a constant value for $I_{ep}$, to help comparisons with Fig.~\ref{fig:Oscillation-period-in}.
One can see that the range of temperatures spanned within the oscillations
is compatible with the values found at constant temperature.

This example shows that oscillations can be expected within a detailed
model using a more sophisticated thermal balance. However, it is not
yet possible to determine the precise domain of parameters involved.

\subsection{$(O/P)$\label{subsec:OsP_ratio}}

Even in steady-state, $(O/P)$ is not equal to its
ETL value. Destruction of $\mathrm{H}_{2}$ by cosmic
rays and reactions with $\mathrm{H}^{+}$ and $\mathrm{H}_{3}^{+}$
are compensated by formation on grains with a different $(O/P)$. However, this
latter process is very poorly understood.
$(O/P)$ is usually supposed to be three
at the formation process on grains, but this is true only if a significant
fraction of the formation enthalpy is used to that end. Some mechanisms
may also lead to a significant transfer of energy to the grain, or
favor a different statistical equilibrium. For example, a strong interaction
with the surface could lead to independent spins for the two protons,
resulting in a $(O/P)$ value of one at formation.

However, our choice is certainly a lower limit, and higher values
are expected, as found by \citet{2006A&A...449..621F}. The thermal
value of $(O/P)$ for the fiducial model at $T=11\,\mathrm{K}$ is
$1.67\,10^{-6}$. We explored the influence of using a higher value
of $(O/P)$ and found that oscillations persist up to $(O/P)=1.65\,10^{-4}$
in this case. We therefore find that oscillations are entirely compatible
with typical values of $(O/P)$.

We checked that the impact of the fraction of ortho-$\mathrm{H}_{2}$
on the occurrence of oscillations is entirely due to
$\mathrm{N}^{+}+\mathrm{H}_{2}\rightarrow\mathrm{NH}^{+}+\mathrm{H}$.
The other $(O/P)$-sensitive reaction, namely
$\mathrm{H_{2}D}^{+}+\mathrm{H}_{2}\rightarrow\mathrm{H}_{3}^{+}+\mathrm{H}$,
has no impact on whether or not oscillations occur.

\subsection{Grain chemistry}

The present study is restricted to pure gas-phase chemistry. However,
recombination of atomic ions on the grain surfaces reduces the ionization
fraction as emphasized by \citet{1995A&A...296..779S} in their discussion
of the effects of grains on bistability properties. We recall that
\citet{1995A&A...302..870L} showed that the bistability domain is
shifted, but not suppressed, when only neutralization on grains is
introduced.

Following the treatment reported in \citet{1995A&A...302..870L} and
summarized in \citet{2017SSRv..212....1C}, we obtain
\begin{equation}
\frac{d\left[\mathrm{X}^{+}\right]}{dt}=-\pi\,a^{2}\,(1+f_{C})\,v_{\mathrm{X^{+}}}\,x_{gr}\,n_{\mathrm{H}}\,\left[\mathrm{X}^{+}\right]\,,
\end{equation}
where $a$ is the grain radius, $v_{\mathrm{X^{+}}}$ is the velocity
of $\mathrm{X}^{+}$, $f_{C}$ is the Coulomb factor ${\displaystyle \frac{2\,e^{2}}{3\,k\,T\,a}}$,
and $x_{gr}$ is the dust-to-gas density ratio. For a typical grain
radius $a$ of $0.1\,\mu$, a dust-to-gas mass ratio $G=10^{-2}$,
a volumic mass density $\rho$ of $3\,\mathrm{g}\,\mathrm{cm}^{-3}$,
and a temperature $T=10\,\mathrm{K}$, $(1+f_{C})=12$, $x_{gr}={\displaystyle \frac{3\times1.4\,m_{\mathrm{H}}\,G}{4\pi\,\rho\,a^{3}}}=1.85\,10^{-12}$
and the equation reduces to:
\begin{equation}
\frac{d\left[\mathrm{X}^{+}\right]}{dt}=-3.2\,10^{-16}\,\frac{1}{\sqrt{M_{\mathrm{X^{+}}}}}\,n_{\mathrm{H}}\,\left[\mathrm{X}^{+}\right]\,,
\end{equation}
where $M_{\mathrm{X^{+}}}$ is the atomic mass of $\mathrm{X^{+}}$
in $\mathrm{amu}$. We find that such a value is compatible with the
oscillatory behavior.

In addition, the neutralization rates are reduced significantly in
low-metallicity regions where the dust-to-gas mass ratio is small,
and in dense cores where coagulation leads to a large mean grain radius
\footnote{The overall dependence of the neutralization rate coefficient on grains
is inversely proportional to $a$, the grain radius.}.

\subsection{Discussion of the oscillatory behavior reported in \citet{2018ApJ...868...41M}}

In a paper devoted to cyanopolyynes and other oxygenated complex organic
molecules in TMC1, \citet{2018ApJ...868...41M} mention oscillatory
solutions for their time-dependent model. Unfortunately, the authors
do not extend their calculations further than two millions years.
Thus, one cannot be sure that these oscillations are maintained over
a large period of time. There are nevertheless striking similarities
between our findings:
\begin{itemize}
\item The oscillatory behavior takes place for specific physical conditions,
corresponding to a large $\mathrm{C/O}$ ratio, low densities, and a large cosmic
ionization rate, which probably would correspond to a chemical HIP
phase, with large abundance of carbon chains.
\item The largest amplitude of the oscillations is obtained for the large
cyanopolyynes whereas we find a similar feature for the highly substituted
deuterated species. These species correspond to the extremities of
successive complex chemical reactions.
\end{itemize}
These authors did not try to further explore such a behavior,
and the origin of these oscillations could be more thoroughly investigated. However,
we note that this model includes grain surface chemistry, which shows
that this is not restricted to a pure gas-phase chemical network.
We also anticipate that the oscillatory behavior concerns all chemical
species, as this is a basic mathematical property of the chemical
equations, as mentioned above. The observed fluctuations of long
chains in TMC1 could come from oscillations triggered with different
initial conditions from place to place. Therefore, variations in abundances
would result from a common origin and common environment.

We also note that these latter authors use initial conditions with all abundances
set to zero except for a single atom or atomic ion for each element (except
for hydrogen). This is not likely to happen in a real cloud that condensates
from diffuse conditions. Hence, their ``early-time'' results are
entirely a consequence of this unfortunate choice and tell us nothing
about the age of the cloud (see Sect.~ \ref{subsec:Initial-conditions-effects}
above). In particular, we can expect that starting from conditions
where most carbon is locked in $\mathrm{CO}$ would quench
the formation of long cyanopolyyne chains entirely if oscillations do not
restore some $\mathrm{C}$ to the gas phase.

\section{Summary and conclusions}

The present study reveals for the first time a well-known feature
of nonlinear effects of chemical kinetics in the context of interstellar
chemistry, namely the occurrence of oscillatory solutions of the chemical
abundances over a restricted but representative physical parameter
space. The oscillations are still present in an isochoric model where
the thermal balance is solved concomitantly with the chemistry. We
obtained different bifurcation diagrams by varying several control
parameters, such as $n_{H}/\zeta$, temperature, and elemental
depletions. The spectacular outcome, in our point of view, lies in
the range of physical parameters where the oscillatory behavior is
obtained, which corresponds to the values derived from interstellar
observations in terms of density, temperature, cosmic ionization rate,
and elemental abundances. In addition, the timescales obtained for the
oscillation periods are between some hundred thousand years and some
million years, which are also within the same order of magnitude as
other important timescale values relevant to the interstellar medium,
such as the free-fall time or the crossing time (\citealt{2016SAAS...43...85K}).

The oscillatory behavior occurs within the so-called HIP phase, corresponding
to electronic fractions which are typically of the order of several
10$^{-7}$, significantly larger than those corresponding to the standard
LIP chemical cloud conditions. These conditions are favored by a relatively
high $\mathrm{C/O}$ elemental ratio which in turn leads to a significant number
of carbon chains, as found in several interstellar conditions.

The triggering mechanism of the oscillations is not precisely identified
but the detailed analysis of the differential behavior of the various
species points to the role of long chains of reactions. In our study,
these are provided by the sequential deuteration processes which end
up with fully deuterated species, such as $\mathrm{CD}_{4}$ and $\mathrm{ND}_{3}$,
and the corresponding molecular ions $\mathrm{CD}_{5}^{+}$ and $\mathrm{ND}_{4}^{+}$.
The amplitude of the oscillations increases dramatically with the
number of substituted deuterium nuclei. We also find that nitrogen
chemistry plays a critical role through the link between carbon and
oxygen.
Finally, we want to emphasize the crucial importance of realistic
initial conditions for deriving sensible conclusions on early-time
behaviors of the chemical abundances.

We can expect that the (magneto-)hydrodynamical behavior of a fluid
subject to chemical oscillations could be quite distinct from that
of a more inert fluid. Various timescales might couple and lead to
unexpected consequences. In particular, we propose that some characteristic
length scales could emerge, which would be a natural explanation for
variations in the star formation rate and in the mass distribution of young
stars in a giant molecular complex, depending on whether or not chemical
oscillations, and the induced oscillation of the cooling function,
are triggered during the collapse.

\appendix

\section{Nitrogen chemistry\label{sec:Nitrogen-chemistry}}

The formation routes of $\mathrm{NH}_{3}$, ..., $\mathrm{ND}_{3}$
from the initial $\mathrm{N}^{+}+\mathrm{H}_{2}$, $\mathrm{N}^{+}+\mathrm{HD}$
and $\mathrm{N}^{+}+\mathrm{D}_{2}$ are displayed in Fig.~\ref{fig:Set-1:-NH3-formation.}.
These reactions have been the subject of a number of experimental
(\citealt{1988JChPh..89.2041M,zymak:13} and references therein) and
theoretical (\citealt{grozdanov:16}) studies. The major issues involved
are due to the mixing of the reactive, rotational, and fine
structure channels of $\mathrm{N}^{+}$, $\mathrm{H}_{2}$, $\mathrm{HD}$,
$\mathrm{NH}^{+}$, $\mathrm{ND}^{+}$, meaning that the reaction proceeds
very differently depending on whether $\mathrm{H}_{2}$ is in its $J=0$, para,
or $J=1$, ortho level, as first noticed by \citet{1991A&A...242..235L}.
Nevertheless, the actual value of the reaction rate coefficient is still uncertain.
The present study was performed by including the experimental
values of \citet{1988JChPh..89.2041M} and following the analytic
prescription of \citet{dislaire:12}, who derived two different analytic
expression for the reactions with para and ortho-$\mathrm{H}_{2}$.
$(O/P)$ is computed from the
gas temperature in our study as we did not try to separate the full
ortho-to-para species as performed in some other studies (\citealt{2018MNRAS.477.4454H,2015A&A...581A.122S}).
The fine tuning of this parameter would allow also to monitor the
deuteration via the $\mathrm{H_{2}D}^{+}+\mathrm{H}_{2}$ reaction
which also critically depends on $(O/P)$
as the endothermicity of the reaction ($192\,\mathrm{K}$) is of the
same order of magnitude as the energy between the $J=1$ and $J=0$
levels of $\mathrm{H}_{2}$ ($170.5\,\mathrm{K}$).

\begin{figure*}
\centering\includegraphics[width=0.8\textwidth]{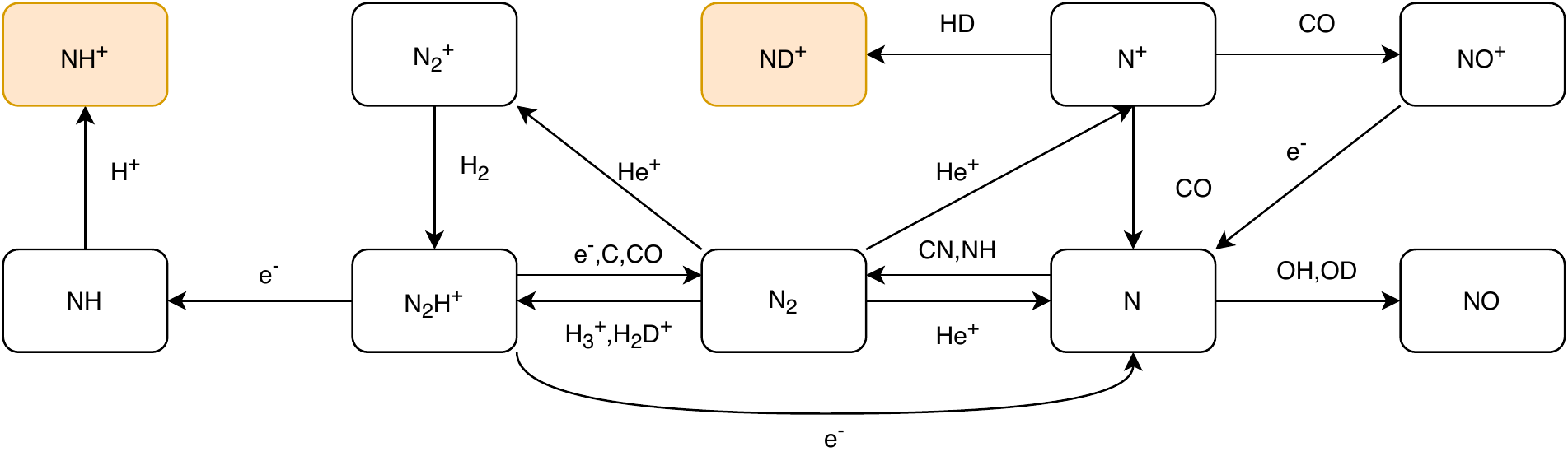}

\caption{Different formation channels of $\mathrm{NH}^{+}$
and $\mathrm{ND}^{+}$.\label{fig:Set-2:-N2-Formation}}

\end{figure*}

As the oscillations take place for low values of the ortho-to-para
ratio of $\mathrm{H}_{2}$, an alternative more efficient formation
route of $\mathrm{NH}^{+}$ in our conditions is in fact obtained
though the charge exchange reaction of $\mathrm{H}^{+}$ with $\mathrm{NH}$
\footnote{We note again a signature of the HIP chemical phase.}. The
corresponding chemical network is displayed in Fig.~\ref{fig:Set-2:-N2-Formation}.

$\mathrm{NH}$ itself is formed through the dissociative recombination
of $\mathrm{N_{2}H}^{+}$. We notice at this point also the crucial
importance of the branching ratio of this reaction, which has been
the object of several contradicting experiments. The present value
is taken as $0.10$, following the latest experimental value of \citet{2020JChPh.152b4301S}.

On the other hand, \citet{1988JChPh..89.2041M} found a relatively
rapid rate coefficient for the $\mathrm{N^{+}}+\mathrm{HD}$ reaction
($3.17\,10^{-10}\,\mathrm{cm}^{3}\,\mathrm{s}^{-1}$), which was used
in the present study. The formation of deuterated ammonia
then follows the reactions starting from $\mathrm{N}^{+}+\mathrm{HD}$ as
shown in Figs.~\ref{fig:Set-1:-NH3-formation.} and \ref{fig:Set-2:-N2-Formation}.

\begin{figure}

\centering\includegraphics[width=0.9\columnwidth]{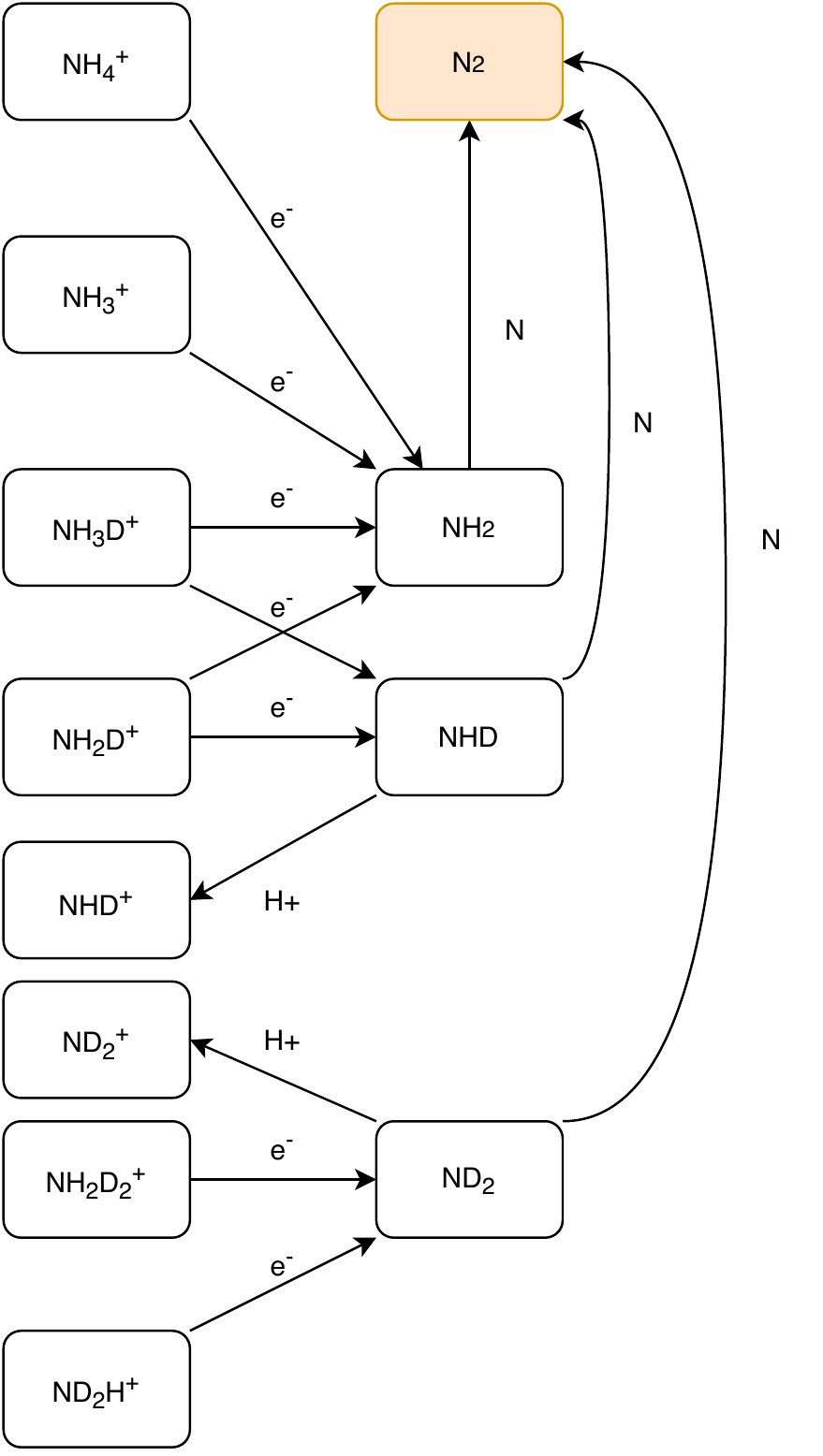}

\caption{Cycle between $\mathrm{NH}_{2}$ and its deuterated
isotopologs and molecular nitrogen.\label{fig:Set-3:-NH-recycling}}

\end{figure}

We also briefly mention $\mathrm{NH}_{2}$ and its deuterated
isotopologs which are obtained from the dissociative recombination
of different protonated molecular ions, as displayed in Fig.~\ref{fig:Set-3:-NH-recycling},
and lead back to $\mathrm{N}_{2}$ via reactions with atomic nitrogen.
This reaction is specifically introduced in \citet{2013arXiv1310.4350W}.

\section{Spectacular critical slow-down\label{sec:Spectacular-critical-slow}}

\begin{figure}
\centering\includegraphics[width=1\columnwidth]{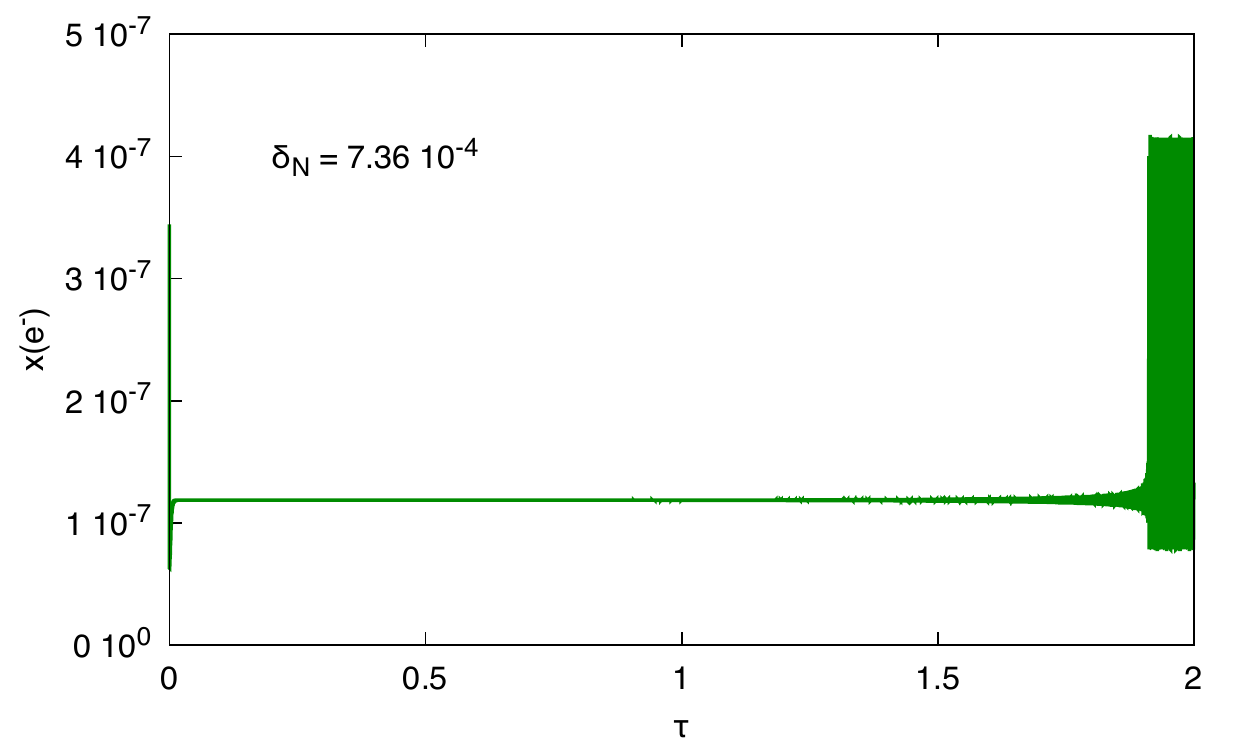}

\caption{Critical slow-down for initial conditions approaching an unstable
fixed point.\label{fig:Critical-slow-down-1}}
\end{figure}

\begin{figure}
\centering\includegraphics[width=1\columnwidth]{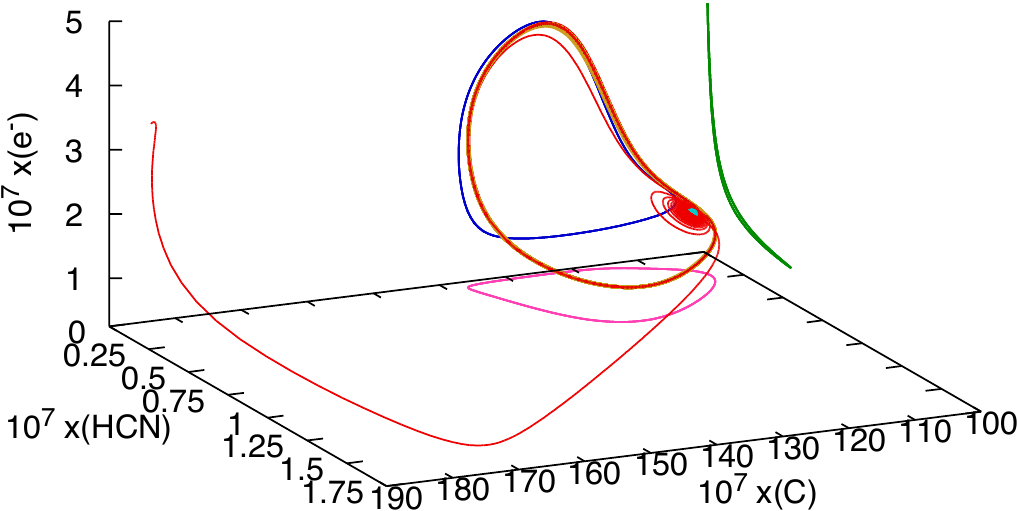}

\caption{Phase space trajectory of the simulation of Fig.~\ref{fig:Critical-slow-down-1}.
The projection of the final limit cycle on each of the three 2D plans
is shown for clarity.\label{fig:Critical-slow-down-Phase}}
\end{figure}

While investigating these bifurcations, we found a nice illustration
of the possible pitfalls offered by systems with multiple possible
asymptotic behaviors (i.e., stable and/or unstable fixed points and
limit cycles). Figure~\ref{fig:Critical-slow-down-1} shows the time
evolution of the ionization degree for $\delta_{\mathrm{N}}=7.36\,10^{-4}$
(which happens, by chance, to be close to the upper bifurcation point).
For more than $1000$ periods, the system seems to have stabilized
at a constant value, which would be the signature of a stable fixed
point. Subsequently, some small fluctuations set in, and the evolution switches
quite abruptly towards large oscillations. This behavior is hard to
interpret if one only looks at the time evolution of one or another
variable; but it becomes much clearer in phase space. Figure~\ref{fig:Critical-slow-down-Phase}
shows one of the many possible projections of the phase-space trajectory.
The initial conditions are on the far left of the figure. The trajectory
relaxes very quickly towards the unstable fixed point (seen in light
blue), close to one of the (many) incoming heteroclinic orbits. However,
once in the immediate vicinity of the fixed point, the influence of
the outgoing heteroclinic orbits associated with positive eigenvalues
of the Jacobian matrix begins to be felt. The trajectory spirals around
the fixed point, with an exponentially increasing size. It goes almost
unnoticed for quite a while, but eventually the trajectory leaves
the vicinity of this fixed point and converges towards the stable
limit cycle, giving rise to oscillations. Starting from different
initial conditions, the trajectory can converge directly towards
this limit cycle within a single period.

\begin{acknowledgements}
This work was supported by the Programme National \textquotedblleft Physique
et Chimie du Milieu Interstellaire\textquotedblright{} (PCMI) of CNRS/INSU
with INC/INP co-funded by CEA and CNES.
\end{acknowledgements}

\bibliographystyle{aa}
\bibliography{Osc}

\end{document}